\documentclass[preprint,12pt]{elsarticle}
\usepackage[top=1in, bottom=1in, left=1in, right=1in]{geometry}

\usepackage{bm}
\usepackage{amsmath}
\usepackage{amssymb}
\usepackage{setspace}
\usepackage{bigints}
\usepackage{array}
\usepackage{multirow}
\usepackage{colortbl}
\usepackage{graphicx}
\usepackage{subfigure}
\usepackage[table,xcdraw]{xcolor}
\usepackage[outdir=./]{epstopdf}
\usepackage{booktabs}
\usepackage{tabu}
\usepackage{bbm}
\usepackage{float}
\usepackage[singlelinecheck=false]{caption}
\usepackage[utf8]{inputenc}
\usepackage[english]{babel}
\usepackage{natbib}
\usepackage{bookmark}
\usepackage{mathrsfs}
\usepackage{color}
\newcommand{\comment}[1]{}
\usepackage{amssymb}
\usepackage{xfrac}
\usepackage{lscape}
\usepackage{pdflscape}
\usepackage{algorithm}
\usepackage{algpseudocode}
\usepackage{xspace}
\usepackage{booktabs}
\usepackage{eucal}

\usepackage[utf8]{inputenc}
\usepackage{tabu}
\usepackage{xcolor}

\begin{document}
\begin{spacing}{1.25}
\begin{frontmatter}


\title{Robust data-driven approach for predicting the configurational energy of high entropy alloys
 \footnote{\footnotesize{This manuscript has been co-authored by UT-Battelle, LLC, under contract DE-AC05-00OR22725 with the US Department of Energy (DOE). The US Government retains and the publisher, by accepting the article for publication, acknowledges that the US government retains a nonexclusive, paid-up, irrevocable, worldwide license to publish or reproduce the published form of this manuscript, or allow others to do so, for US government purposes. DOE will provide public access to these results of federally sponsored research in accordance with the DOE Public Access Plan (http://energy.gov/downloads/doe-public-access-plan).}}}



\author{Jiaxin Zhang\corref{cor2}$^\dagger$ \footnote{\footnotesize{ $^\dagger$ These two authors contributed equally to this work}}}
\ead{jiaxin.zhanguq@gmail.com}
\address{Center for Computational Sciences, Oak Ridge National Laboratory}

\author{Xianglin Liu\corref{cor1}$^\dagger$}
\ead{xianglinliu01@gmail.com}
\address{Materials Science and Technology Division, Oak Ridge National Laboratory}

\author{Sirui Bi}
\address{Department of Civil Engineering, Johns Hopkins University}
\author{Junqi Yin}
\address{Center for Computational Sciences, Oak Ridge National Laboratory}

\author{Guannan Zhang}
\address{Computer Science and Mathematics Division, Oak Ridge National Laboratory}

\author{Markus Eisenbach}
\address{Center for Computational Sciences, Oak Ridge National Laboratory}

\begin{abstract}
High entropy alloys (HEAs) have been increasingly attractive as promising next-generation materials due to their various excellent properties. It's necessary to essentially characterize the degree of chemical ordering and identify order-disorder transitions through efficient simulation and modeling of thermodynamics. In this study, a robust data-driven framework based on Bayesian approaches is proposed and demonstrated on the accurate and efficient prediction of configurational energy of high entropy alloys. The proposed effective pair interaction (EPI) model with ensemble sampling is used to map the configuration and its corresponding energy. Given limited data calculated by first-principles calculations, Bayesian regularized regression not only offers an accurate and stable prediction but also effectively quantifies the uncertainties associated with EPI parameters. Compared with the arbitrary determination of model complexity, we further conduct a physical feature selection to identify the truncation of coordination shells in EPI model using Bayesian information criterion. The results achieve efficient and robust performance in predicting the configurational energy, particularly given small data. The developed methodology is applied to study a series of refractory HEAs, i.e. NbMoTaW, NbMoTaWV and NbMoTaWTi where it is demonstrated how dataset size affects the confidence we can place in statistical estimates of configurational energy when data are sparse. 

%

\end{abstract}


\end{frontmatter}


\section{Introduction}
As one of the typical multicomponent alloys, high entropy alloys (HEAs) consisting of four or more principal elements have been widely studied due to their exceptional mechanical properties \cite{yeh2004nanostructured, senkov2011mechanical, zhang2014microstructures, li2016metastable}. The increased number of elements expand the possible combinations and potential candidates for discovering next-generation materials with enhanced properties \cite{tsai2014high,miracle2017critical,gao2016high}. Typically, the material properties are inherently linked to the actual state of chemical ordering, much efforts have been therefore devoted to analyze the degree of chemical ordering and to identify the order-disorder phase transitions \cite{widom2014hybrid, gao2016high, murty2019high, eisenbach2019first}. Due to expensive time costs in experimental research, computational simulations, typically first-principles calculations are playing an increasingly central role in the investigation of various properties of HEAs \cite{widom2018modeling, gao2017computational, ikeda2019ab}. 

First-principles density functional theory (DFT) methods have established as a powerful and reliable tool in computational material science and have enabled critical advancements in materials properties and performance discovery \cite{ye2016high, ma2015ab}. With the increasing numerical efficiency and growing computing power (parallel and GPU computing), it is still difficult to address the challenge of DFT calculations in relatively large supercells (thousands of atoms) and intensive sampling (huge number of configurations) \cite{huang2017construction}. To characterize the order-disorder phase transition, a straightforward way is to combine the DFT method with Monte Carlo simulations. However, this ``brute-force" method is so computationally intensive that it is often impractical, even for a simple example, 250-atom CuZn alloy \cite{khan2016density}. Consequently, it is quite necessary to establish an approximate configurational energy model fitted to DFT data and feed this accurate enough and efficient ``surrogate" model into Monte Carlo simulations for modeling thermodynamics and order-disorder phase transitions. 

A typical and effective strategy is cluster expansion (CE) method \cite{kikuchi1951theory, sanchez1984generalized, van2002automating}, which uses discrete sum representation of material properties, for example, configurational energies, in terms of lattice site configuration and site effective cluster interactions (ECIs), such as site pairs, triplets, quadruplets and the other high order interactions. The fundamental challenge in constructing CE model is to determine the ECIs in an efficient and robust way through a fit to reference configuration. Commonly used fitting algorithm is to minimize the overall difference between the CE fitted energy and DFT calculated energies with respect to different input configurations. Practically, CE has to be truncated and many advanced methods have been proposed to aid in the efficient and accurate performance of a truncated CE. These methods include as compressive sensing\cite{nelson2013compressive}, Bayesian method \cite{mueller2009bayesian}, cluster basis set selection\cite{blum2005using, hart2005evolutionary, seko2009cluster}, machine learning \cite{natarajan2018machine} and regularization \cite{chang2019clease, aangqvist2019icet}. Although the CE is effective Hamiltonian and can be potentially combined with Monte Carlo simulation to account for order-disorder phase transitions and chemical short-range order, its application to multicomponent systems is still intractable due to rapid increased combinatorial number of interatomic interactions concerning chemical elements \cite{jiang2016efficient, seko2009cluster}. Therefore, fitting a CE for multicomponent alloys, i.e. HEAs, becomes extremely difficult \cite{widom2018modeling}. 

The recent development of machine learning presents exciting opportunities and challenges to various scientific fields \cite{jordan2015machine, lecun2015deep}. Benefiting from advanced learning algorithms and large databases using high-throughput computations, machine learning has been widely applied to materials research and discovery\cite{butler2018machine, mueller2016machine, sanchez2018inverse}. Some examples of successful applications include discovering complex materials behavior\cite{kim2016organized, raccuglia2016machine}, accurate prediction of phase transitions and prediction\cite{carrasquilla2017machine, huang2019machine, kostiuchenko2019impact}, accelerated material design and prediction of material properties\cite{fujimura2013accelerated, pilania2013accelerating, ward2016general}, modeling of various physical quantities, for instance, interatomic potentials \cite{dragoni2018achieving, chmiela2017machine, deringer2017machine} and atomic forces\cite{chmiela2017machine, li2015molecular}. Compared to successful applications in other fields, few studies have been conducted in the context of machine learning for the modeling of thermodynamics of HEAs. This is because the inherent challenges originated from the extremely large configuration space associated with the multicomponent alloys. In many cases, only small dataset is drawn from expensive DFT calculations due to limited computational cost, it therefore gives rise to the issue of uncertainty quantification in model inference and certified predictions \cite{aldegunde2016quantifying, kristensen2014bayesian, zhang2018quantification}. The learned model also faces additional challenges to capture the underlying physics with important features and cover the overall configurational space in an accurate and robust scheme \cite{liu2019machine}. 

In this work, we develop an efficient and robust Bayesian framework by fitting an accurate, feature-selected efficient Hamiltonian which is employed in subsequent Monte Carlo simulations for modeling the thermodynamics and order-disorder phase transitions. Bayesian regularized regression is employed to deal with the unstable prediction due to sparse data, on the other hand, to effectively quantify the uncertainties associated with the EPI parameters. To investigate the impact of model complexity, we conduct physical feature selection using Bayesian information criterion that allows for effective truncation of the coordination shells given a specific dataset. We demonstrate the accuracy and reliability of prediction with feature selection are significantly higher than the prediction with an arbitrary truncation, specifically when data are limited. This first section of this paper presents a brief overview of classical cluster expansion and the proposed effective pair interaction model \cite{liu2019machine}. Then we propose a robust data-driven framework which consists of Bayesian regularized regression and Bayesian feature selection that enables to effectively reduce the model complexity that is very difficult to conventional CE model in HEAs. In the second section, we apply the proposed robust algorithm to three refractory HEAs, i.e. NbMoTaW, NbMoTaWV and NbMoTaWTi. The ensemble random sampling performs well in predicting of configurational energy compared with the approach that uses only single supercell. Due to the limitation of dataset size, we systematically present the uncertainty quantification and correlation analysis of model parameters in terms of the Bayesian framework. Moreover, the effect of physical feature selection is carefully investigated for these three HEAs. Finally, the conclusions of the current work are summarized.


\section{Theory and algorithm}

\subsection{Re-visited cluster expansion method}
cluster expansion (CE) method is widely used approach for thermodynamic simulation of binary alloys due to its versatility and simplicity. Specifically, a binary alloy $XY$ of $N$ sites can be represented as a vector of occupation configurations, $\bm{\sigma} = \left\{ \sigma_1, ... , \sigma_k, ... , \sigma_N \right\}$ where `spin' variable $\sigma_k$ takes a value of $-1$ or $+1$ depending on the occupant (atom X or Y) of site $k$. A property of this binary crystal that depends on $\bm{\sigma}$ can be formulated as a polynomial expansion in terms of occupation configurations 
\begin{equation}
F(\bm{\sigma}) = NV_0 + \sum_{\beta} V_{\beta}^{(n)}\Phi_{\beta}^{(n)}(\bm{\sigma}) \label{eq:ce1}
\end{equation}
where the expansion coefficients $V_{\beta}^{(n)}$ are called the effective cluster interactions (ECIs) which are independent of the configurations and often determined by the crystal structure and chemistry of the binary alloy, $V_0$ is a constant that represents the empty cluster and $\Phi_{\beta}^{(n)}(\bm{\sigma})$ is the $n$-site cluster function, defined as the product of basis function ${ \Theta_{\beta_k} (\sigma_k) }$, which is given by:
\begin{equation}
\Phi_{\beta}^{(n)}(\bm{\sigma}) = \prod _{}^{ }{ \Theta_{\beta_k} (\sigma_k) } \label{eq:ce2}
\end{equation} 
Note that the cluster functions in Eq. \eqref{eq:ce2} form a complete orthonormal basis on the configuration space $\bm{\sigma}$. When all possible cluster functions are considered in CE model, Eq. \eqref{eq:ce1} is an exact expression. However, a more practical way for CE is a truncated summation over finite number of cluster functions considering the many sites that are far apart are usually negligible. Typically, the energy is primarily determined by short-range interactions, it is therefore natural to represent the energy as a summation of interactions whose strength diminished with increasing range, which is given by:
\begin{equation}
E(\bm{\sigma}) = \sum_{i} V_i { \Phi_{i} (\sigma_i) } + \sum_{ij} V_{ij} { \Phi_{ij} (\sigma_i, \sigma_j) } + \sum_{ijk} V_{ijk} { \Phi_{ijk} (\sigma_i, \sigma_j, \sigma_k)  + \cdots} \label{eq:ce3}
\end{equation}
where the $V_i$, $ V_{ij} $ and $ V_{ijk} $ represent the interaction strength of point clusters, pair clusters and triplet clusters, and can be determined by the DFT calculated total energies of different configurations in a variety of supercells. One can therefore utilize CE as an efficient Hamiltonian in Monte Carlo simulation to reveal order-disorder phase transitions. 

However, it is often a challenging task to fit the ECIs of CE in multicomponent crystalline solids because of the number of terms scales as $N^M$ for an $M$-body terms for $N$ species. The series need many terms and the number of terms grows rapidly with the diameter of the cluster. As additional terms are added, the series coefficients may converge poorly given limited number of configurations. 
 
\subsection{Effective pair interaction model}
\begin{figure}[!ht]   
  \centering
  {\includegraphics[width=0.8\textwidth]{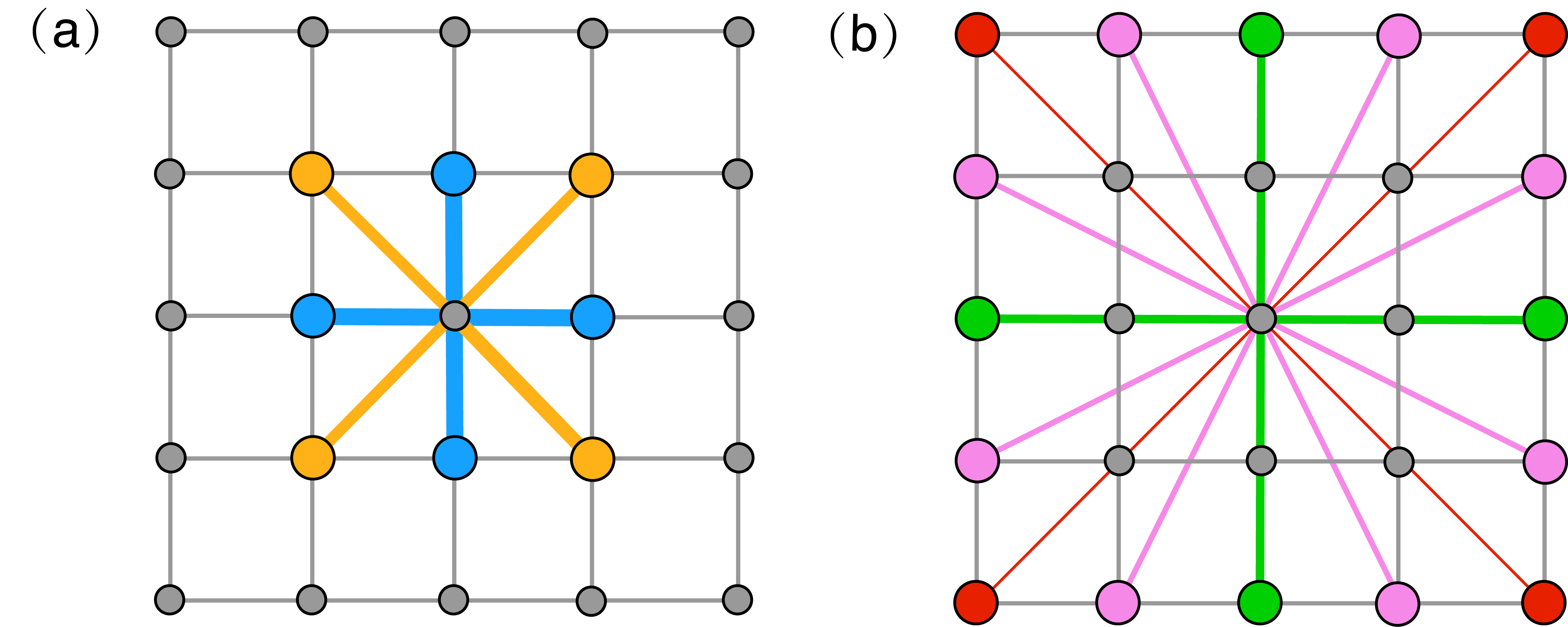}} \quad
  \caption{Square lattice with effective pair interaction highlighted. (a) the nearest-neighbor pair is marked in blue, while the next nearest-neighbor pair is marked in yellow; (b) the pair marked in green, pink and red correspond to the 3rd, 4th and 5th neighbor respectively. Equivalent interacted pairs (same distance) are marked in the same color. } \label{fig:s2_1}
\end{figure}

Conventional CE method is difficult when applied to HEAs, we herein propose to use an Ising-like model with only effective pair interactions (EPIs) without considering high-order interactions. Fig. \ref{fig:s2_1} shows the prototype square lattice with effective pair interactions. In terms of the pair distance, we define a series of short-range pair interactions, for example, the nearest-neighbor, the next nearest-neighbor and so on. The orbit of a specific pair centered around site $i$ consists of all the equivalent pair interactions. Therefore, the effective Hamiltonian at lattice site $i$ can be expressed as: 

\begin{equation}
H(i) =J_0 + \sum_{j\neq i} J_m^{X(i)Y(j)}c_j \label{eq:epi1}
\end{equation}
where $J_m^{X,Y}$ is the interatomic pair potential between element $X$ and $Y$, $X(i)$ is referred to as element $X$ at site $i$, $m$ is the number of coordination shell separating between $i$ and $j$, $c_j$ is the occupation parameter, and $J_0$ is the concentration dependent part, which can be discarded for a given composition. Summing up the Hamiltonian over all atomic sites yields the total energy, which is given by:
\begin{equation}
\tilde{E}(\bm{\sigma}) =NJ_0 + N\sum_{X, Y, m} J_m^{X, Y}\sigma_{m}^{X,Y} \label{eq:epi2}
\end{equation}
 where $\sigma_m^{X,Y}$ is the percentage of $XY$ bonds in the $m$-th coordination shell. Considering an $n$-component alloy system, the total number of different chemistry bonds in $m$-th shell is $n(n+1)/2$ but there are $n$ constraints from the concentration of each element for a fixed chemical composition, for example, $X$-$X$ and $Y$-$Y$. As a result, the number of independent variables in an $n$-component alloy system is 
 \begin{equation}
 N_m = \frac{n(n+1)}{2} - n = \frac{n(n-1)}{2} \label{eq:epi3}
 \end{equation}
 which consists of the nearest-neighbor short-range order (SRO) parameters that exist at $m$-th shell for an $n$-component alloy. The Warren-Cowley SRO parameters is defined as 
 \begin{equation}
 \alpha_m^{X,Y} = 1 - \frac{P_m^{X|Y}}{c_A} \label{eq:epi4}
 \end{equation}
 where $c_A$ is the concentration of element $X$, and $P_m^{X|Y}$ is the probability of finding element $X$ at the $m$-th neighbor shell of element $Y$. $\alpha_m^{XY}$ is a critical parameter to characterize the different chemical configurations. $\alpha_m^{XY}>0$ means the preference of form $XY$ bonds at the $m$-th shell, $\alpha_m^{XY}<0$ indicates the opposites and $\alpha_m^{XY}=0$ for each $m$ suggests to a completely random system. In fact, there is an effective pair interaction (EPI) corresponding to each SRO parameter. Consequently, Eq. \eqref{eq:epi2} can be further written by
 \begin{equation}
E= N \sum_{X \neq Y, m} J_m^{X, Y}P_m^{X | Y} \label{eq:epi5}
 \end{equation}
 where $P_m^{X | Y}$ is closely related to the SRO parameter in Eq. \eqref{eq:epi4}. For example, for the four-component Nb-Mo-Ta-W refractory HEAs, there are total ten different bonds for each coordination shell but only six independent bonds, including Nb-Mo, Nb-Ta, Nb-W, Mo-Ta, Mo-W and Ta-W. Given a specific configuration of multicomponent HEAs Nb-Mo-Ta-W, it is not difficult to calculate the $P_m^{Nb|Mo}$, $P_m^{Nb|Ta}$, $P_m^{Nb|W}$, $P_m^{Mo|Ta}$, $P_m^{Mo|W}$ and $P_m^{Ta|W}$ at the $m$-th neighbor shell. The corresponding interatomic pair coefficients $V_m^{Nb|Mo}$, $V_m^{Nb|Ta}$, $V_m^{Nb|W}$, $V_m^{Mo|Ta}$, $V_m^{Mo|W}$ and $V_m^{Ta|W}$ at the $m$-th neighbor shell in Eq.\eqref{eq:epi5} can be determined by linear regression using $P_m^{X | Y}$ as the features\cite{liu2019machine}. The cost of building an EPI model comes primarily from the cost of generating the training dataset, which is usually by DFT to which the EPIs are fitted. It therefore gives rise to two critical questions: 1) how to conduct an accurate and robust prediction that minimizes the error of energy for a given training dataset and 2) how to determine the number of physical feature $m$ when data are sparse. These would correspond to cluster selection in which the optimal set of clusters is selected for inclusion in the expansion in a robust way that minimizes the expected prediction error. 
 
\subsection{Bayesian regularized regression}
To obtain a EPI for a specific HEAs one must determine the interatomic pair potential $J_m^{X,Y}$ in Eq. \eqref{eq:epi5}, which can be cast in a matrix form, 
\begin{equation}
\bm{E} = \bm{JP} + \bm{\varepsilon} \label{eq:Bayes0}
\end{equation}
where $\bm{\varepsilon} = (\varepsilon_1,\varepsilon_2, ..., \varepsilon_n)^T$ are independent and identically distributed (i.i.d.) variables that follow $ \bm{\varepsilon} \sim N(0, \sigma^2) $. $\bm{P}$ is a matrix containing the probability quantities of the training data where each element in row $i$ at the $m$-th shell is defined as 
\begin{equation}
\bm{P}_m^{i}= P_m^{X|Y} (\bm{\sigma}_i), \quad i = 1,...,N_{\sigma} \label{eq:Bayes1}
\end{equation}
where $N_{\sigma}$ is the number of training data (configurations). It is necessary to note that there are $N_m = n(n-1)/2$ column vectors for each shell and $\bm{P}$ is therefore a $N_{\sigma} \times N_m$ matrix. $\bm{E}$ is a column vector in which the $i$th element in the physical quantity $E_i$ (for example, total energy) of the configuration $\bm{\sigma}_i$ and $\bm{J}$ is a column vector in which $m$-th shell is $J_m$ which also includes $N_m = n(n-1)/2$ elements.  
Determining the EPIs by solving the linear system given by Eq. \eqref{eq:Bayes1}, it is equivalent to finding the parameter vector $\bm{J}$, which minimizes the residual sum of squared errors (RSS) $ \left\| \bm{JP}-\bm{E}\right\|_2^2$ using ordinary least squares (OLS) method. Typically, OLS method often has low bias but larger variance. A solution can be determined by OLS method, which performs well in the overdetermined system. Due to the large computational cost in DFT calculations, the linear system in Eq. \eqref{eq:Bayes1} is, however, often underdetermined and therefore leads to an ill-posed problem. Another drawback associated with OLS method is the susceptibility to possible overfitting \cite{chang2019clease}, which refers to that the EPIs values are over-tuned to predict physical quantity in training dataset but losing the predictability for the new configurations that are ``unseen" before. Meanwhile, the nearsightedness of physical interaction in CE suggests sparse property for $\bm{J}$ \cite{ nelson2013compressive}. 

Regularization is an effective way to combat overfitting and achieve sparse solutions by adding a regularization term in the form of $\ell_1$ or $\ell_2$ norm. For $\ell_1$ regularization, the optimal EPI values $\hat{\bm{J}}$ can be solved by
\begin{equation}
\hat{\bm{J}} = \underset { \bm{J} }{\arg\min} \left\| \bm{JP}-\bm{E}\right\|_2^2 + \lambda_1 \left\| \bm{J}\right\|_1 \label{eq:Bayes2}
\end{equation}
where $\lambda_1$ is a penalty parameter that determines the amount of regularization. The primary benefit of $\ell_1$ regularization is its promotion of sparsity, which is achieved by feature selection with a set of EPI values set to zero. However, in principle, this shrinkage is performed only based on the correlation of feature but ignoring the inherent physical interactions such that sometimes it incorrectly forces the EPI parameters to zero, consequently leads to a unstable prediction, specifically under the case of sparse training dataset. Instead, this study prefers to use the $\ell_2$ penalty for both fitting and penalization of the EPI coefficients. Thus, the solution becomes 
\begin{equation}
\hat{\bm{J}} = \underset { \bm{J} }{\arg\min} \left\| \bm{JP}-\bm{E}\right\|_2^2 + \lambda_2\left\| \bm{J}\right\|_2^2 \label{eq:Bayes3}
\end{equation}
which is the most popular technique for improving prediction accuracy by shrinking large regression coefficients to reduce overfitting. Unlike $\ell_1$ regularization, $\ell_2$ regularization tends to contain all physical interaction information by only shrinking the size of EPI coefficients rather than set most of them to zero. It therefore gives rise to challenges in optimally determining the $\ell_2$ regularization parameter and physically identifying the important features. Moreover, it is critical to quantify the uncertainties associated with the prediction and EPI coefficients, particularly given lack of data. 

In this paper, we propose a Bayesian view of regression with feature selection to address these challenges. Bayesian regression assumes the parameters $\bm{J}$ and $\sigma^2$ in Eq. \eqref{eq:Bayes0} to be the random variables, therefore the likelihood function can be written as:
\begin{equation}
p(\bm{E} | \bm{J}, \bm{P}, \sigma^2) \propto (\sigma^2)^{-n/2}\exp \left( -\frac{1}{2\sigma^2} (\bm{E}-\bm{JP})^T(\bm{E}-\bm{JP}) \right) \label{eq:Bayes4}
\end{equation}

Bayesian regression can be also used to take $\ell_2$ regularization into consideration in the estimation procedure. Instead of identifying the optimal $\lambda_2$ in a hard sense, Bayesian regression treats the regularization parameter $\lambda_2$ as a random variable that can be estimated via the training data. This can be achieved by introducing hierarchical model with hyper-parameters of the model. In the Bayesian setting, the target total energy $E$ is assumed to be Gaussian distribution, which is given by:
\begin{equation}
p(\bm{E} | \bm{J}, \bm{P}, \lambda_2) = \mathcal{N}(\bm{E} | \bm{JP}, \lambda_2)
\end{equation}
and the prior for the EPI coefficient $\bm{J}$ is given by a Gaussian distribution
\begin{equation}
p(\bm{J} | \xi ) = \mathcal{N}(\bm{J} | 0, \xi ^{-1} \bm{I}_p)
\end{equation}
Consequently, the $\ell_2$ regularization in Eq. \eqref{eq:Bayes3} is equivalent to finding a maximum a posterior (MAP) estimation \cite{zhang2019efficient} given a Gaussian prior over $\bm{J}$ with precision $\xi^{-1}$. Typically, a MAP estimation of the posterior distribution is obtained by Markov Chain Monte Carlo (MCMC) algorithm, which is often computationally intensive and difficult to converge for high dimensional problem. In this work, we consider a so-called conjugate prior for which the posterior distribution can be derived analytically and thus be more efficient. To this end, the priors over $\lambda_2$ and $\xi$ are selected to be gamma distribution 
\begin{equation}
\lambda_2 \sim \mathcal{G}(\alpha_1, \alpha_2), \quad \xi \sim \mathcal{G}(\beta_1, \beta_2)
\end{equation}
where $\alpha_1, \alpha_2, \beta_1$ and $\beta_2$ are the hyperparameters of the gamma priors over $\lambda_2$ and $\xi$. We here select $\alpha_1 = \alpha_2 = \beta_1 = \beta_2 = 10^{-8}$ to be non-informative priors. All three random variables $\bm{J}$, $\lambda_2$ and $\xi$ are estimated jointly using maximum likelihood estimate during the fit of the regression model. Note that Bayesian regularized regression performs more robust to ill-posed problems. 

\subsection{Bayesian feature selection}
To construct the effective Hamiltonian in HEAs, EPI model is employed to identify the coordination shells as the essential physical features. For each shell, there are $N_m=n(n-1)/2$ independent sub-features that is determined by the number of element species, for instance, $N_m=6$ for four-component Nb-Mo-Ta-W HEAs and $N_m=10$ for five-component Nb-Mo-Ta-W-V HEAs. Liu and Zhang \cite{liu2019machine} have shown that the first two shells associated with nearest neighbor pair and the next nearest neighbor pair have a more significant impact on the accuracy of prediction but long-range pair interactions with a larger $m$ perform weak influence. In practice, the truncation of coordination shells $m$, needs to be carefully examined and determined for each specific material. In this work, Bayesian model selection method is applied to identify a ``better" model complexity among a finite set of candidate EPI models.  

Bayesian information criterion (BIC) is widely used for feature selection and measure the efficiency of the parameterized model in terms of predicting the data, which is defined as
\begin{equation}
\textup{BIC} =-2\log(\hat{L})+k \log(n_d) \label{eq:BIC1}
\end{equation}
where $\hat{L}$ is the maximized value of the model likelihood function, i.e. $\hat{L} = p(\bm{d} | \bm{J}^*, M)$, where $\bm{J}^*$ are the parameter values that maximize the likelihood function. $n_d$ is the number of observed data $\bm{d}$, $k$ is the number of parameters (features) of the model $M$. BIC is derived by an efficient approximation using Laplace's approach to approximate the evidence $p(\bm{d}|M)$, which is defined by Bayesian inference \cite{zhang2018effect}
\begin{equation}
p(\bm{J} | \bm{d}, M) = \frac{p(\bm{d}|\bm{J}, M) p(\bm{J} | M)}{p(\bm{d}|M)} = \frac{p(\bm{d}|\bm{J}, M) p(\bm{J} | M)}{\int p(\bm{d}|\bm{J}, M) p(\bm{J} | M) d\bm{J}} \label{eq:BIC2}
\end{equation}

BIC can be extended for linear regression under the assumption that the model errors $ \bm{\varepsilon}$ are i.i.d. random variables that follow Gaussian distribution $N(0, \sigma^2)$. The likelihood of $\bm{\varepsilon}$ can be written as 
\begin{equation}
L = \prod _{ i=1 }^{n_d }{ \frac{1}{\sigma \sqrt{2\pi}} \exp \left( -\frac{(\bm{E}_i -\bm{J} \bm{P}_i )^2}{2\sigma^2} \right) } = \frac{1}{(2\pi)^{\frac{n_d}{2}} \sigma^{n_d}} \exp\left(-\frac{\sum_{i=1}^{n_d}(\bm{E}_i -\bm{J} \bm{P}_i )^2}{2\sigma^2} \right) \label{eq:BIC3}
\end{equation}
Note that $\sum_{i=1}^{n_d}(\bm{E}_i -\bm{J} \bm{P}_i )^2 = \left\| \bm{JP}-\bm{E}\right\|_2^2$ is the RSS of $\bm{\varepsilon}$. Taking the derivative of $L$ with respect to $\sigma$ and equate to zero yields the maximized value of $L$ and the corresponding log of $\hat{L}$ is $\log(\hat{L}) = -n_d/2\log(\textup{RSS}/n_d)$. As a result, the BIC in terms of RSS is given by 
\begin{equation}
\textup{BIC}_{\textup{RSS}} =n_d\log\left(\frac{\textup{RSS}}{n_d} \right)+k \log(n_d) \label{eq:BIC4}
\end{equation}
This $\textup{BIC}_{\textup{RSS}}$ can be used to identify an appropriate number of physical feature, the coordination shell in EPI model, according to the intrinsic complexity present in a particular dataset. When selecting from a set of candidate models with various number of features, the one with lowest $\textup{BIC}_{\textup{RSS}}$ value is preferred. 

\subsection{Robust data-driven algorithm procedure}

\begin{figure}[!ht]   
  \centering
  {\includegraphics[width=0.6\textwidth]{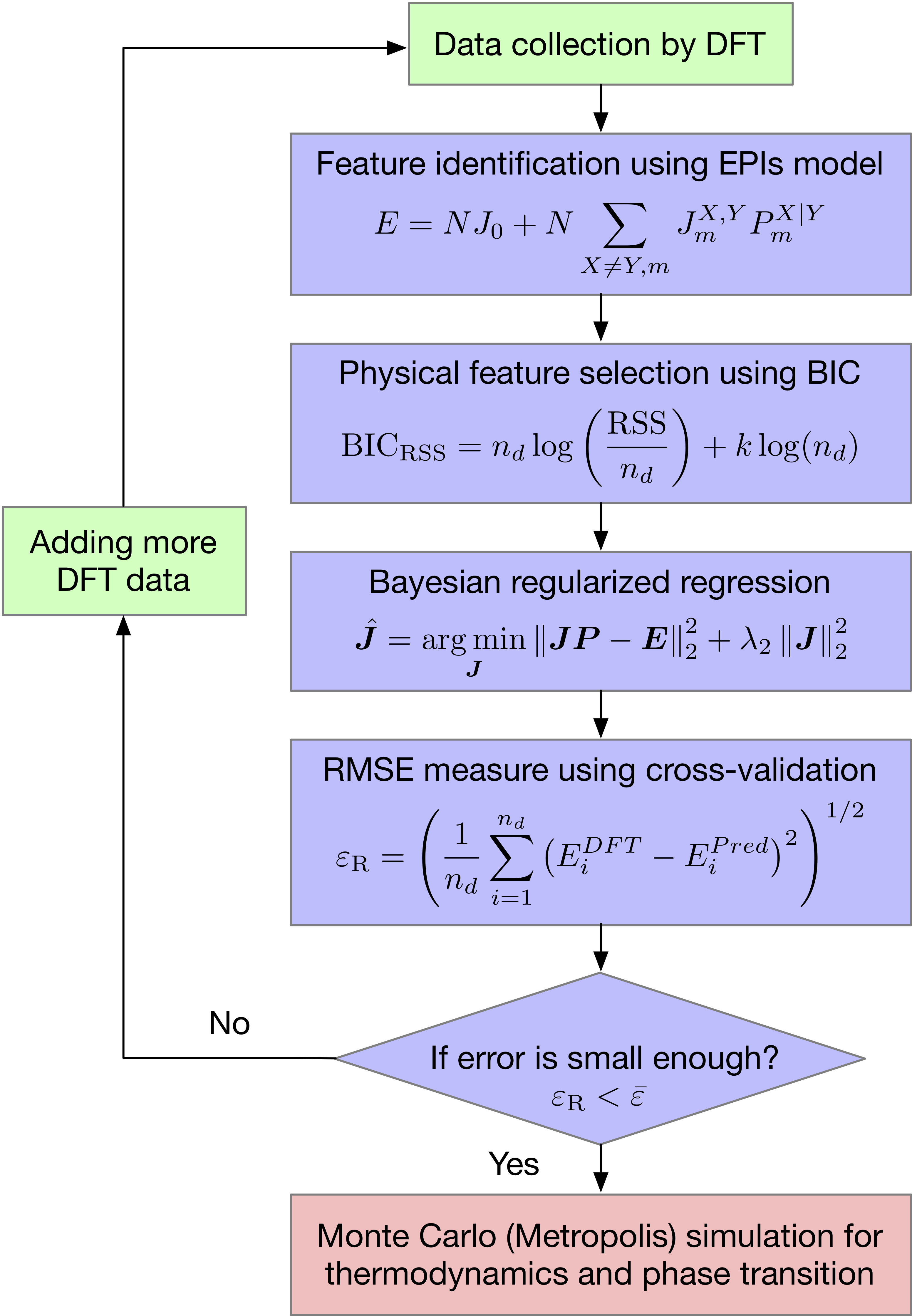}} \quad
  \caption{Flowchart of robust data-driven algorithm using Bayesian framework} \label{fig:flowchart}
\end{figure}

With the constituents outlined in the previous sections, the proposed methodology is summarized here and a flowchart is provided in Fig. \ref{fig:flowchart}. 

\begin{itemize}
\item Step 1: \emph{Data collection} - An ensemble sampling strategy is used to combine the DFT data calculated with different sizes of the supercell. This benefits from incorporating different long-range order and short-range order dataset. 
\item Step 2: \emph{Feature identification} - Given a set of random configurations of $n$-component alloy, the first step is to determine the $N_m = n(n-1)/2$ independent pair interactions according to the EPI model. Then the probabilities $P_m^{X|Y}$, as physical features, defined in Eq. \eqref{eq:epi4}, are carefully calculated for the $m$-th coordination shell. 

\item Step 3: \emph{Feature selection} - To achieve an accurate prediction and reduce overfitting, feature selection using BIC in Eq. \eqref{eq:BIC4} is employed here to determine the truncated number of coordination shells $m$ for a specific HEAs. Note that each shell also includes $N_m = n(n-1)/2$ sub-features, which are either fully retained or truncated as a whole in the $m$-th shell. 

\item Step 4: \emph{Bayesian regularized regression} - Bayesian regression with $\ell_2$ regularization performs a robust prediction for the configurational energy $\bm{E}_i$ given a specific configuration $\bm{\sigma}_i$. Under the assumption of Gaussian distribution with conjugate prior, the uncertainty of the EPIs parameters are efficiently quantified, particularly given limited DFT data due to prohibitive computational cost. 

\item Step 5: \emph{Error evaluation and model update} - The performance of predicted model can be assessed by k-fold cross validation (k=5) with the root-mean-square error (RMSE) metric $\varepsilon_{\textup{R}}$, which is defined as:
\begin{equation}
\varepsilon_{\textup{R}} = \left(\frac{1}{n_d}\sum_{i=1}^{n_d}\left(E_i^{DFT}-E_i^{Pred} \right)^2\right)^{1/2}
\end{equation}
where $E^{DFT}$ is the true value of energy by DFT and $E^{Pred}$ is the predicted energy using the proposed methodology. If the RMSE metric $\varepsilon_{\textup{R}}$ is smaller than a specific threshold $\bar { \varepsilon} $, for example, $\bar { \varepsilon}$ = 1 meV, the predictive accuracy is acceptable, otherwise additional DFT data are required and back to Step 1 to update the algorithm and further improve the performance. 

\item Step 6: \emph{Monte Carlo calculations of thermodynamics} - If the prediction is accurate and stable enough, we can feed this efficient fitted model as a surrogate into Monte Carlo simulation for modeling thermodynamics and order-disorder phase transitions. But this step is not the focus of this paper and the interested reader can find more discussions in the recent review literature \cite{widom2018modeling, eisenbach2019first}

\end{itemize}

\section{Results and discussions}
\subsection{Prediction of configurational energy} 
\begin{figure}[!ht]   
  \centering
  {\includegraphics[width=0.9\textwidth]{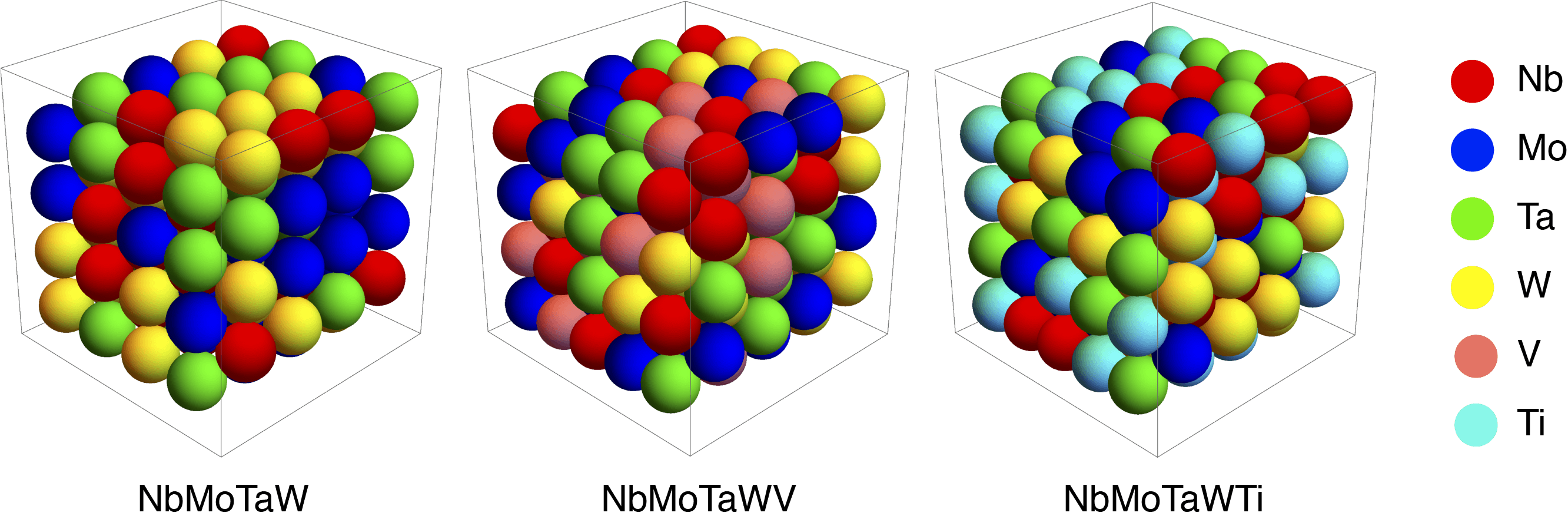}} \quad
  \caption{ Bcc supercells of refractory HEAs. (a) NbMoTaW with 128 atoms, (b) NbMoTaWV with 160 atoms and (c) NbMoTaWTi with 160 atoms} \label{fig:s3_0}
\end{figure}

\begin{figure}[!ht]   
  \centering
  {\includegraphics[width=1\textwidth]{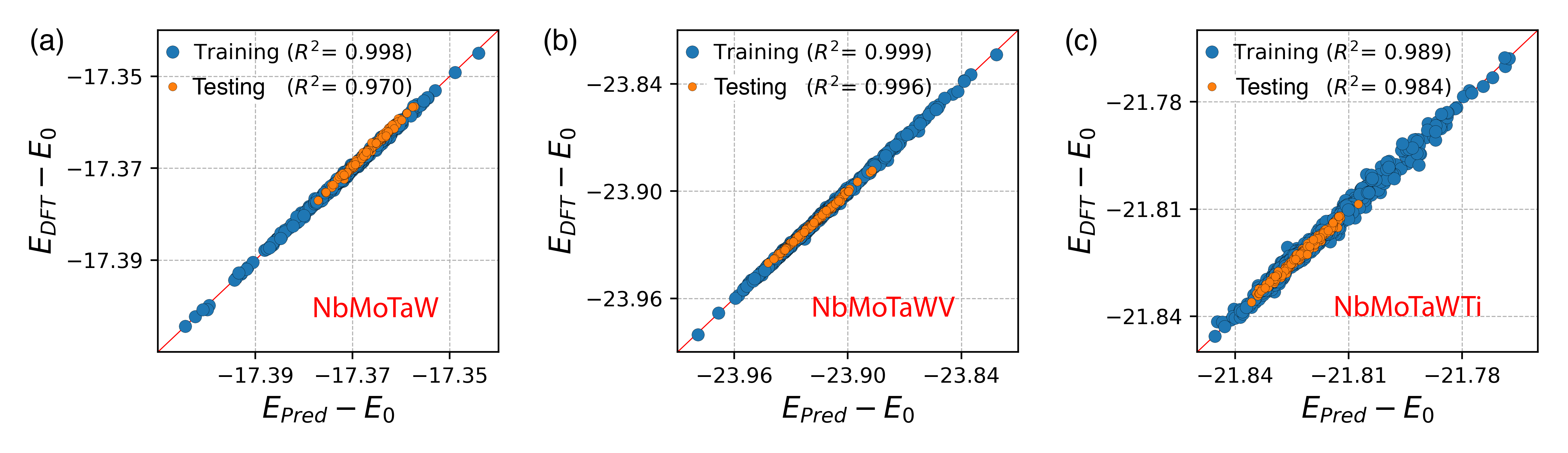}} \quad
  \caption{Comparison of DFT calculated energy with predicted energy using Bayesian regularized regression for (a) NbMoTaW, (b) NbMoTaWV and (c) NbMoTaWTi } \label{fig:s3_1}
\end{figure}

In this work, we systematically investigate three HEAs, including NbMoTaW, NbMoTaWV and NbMoTaWTi. The locally self-consistent multiple scattering (LSMS) method \cite{wang1995order} is used here for the calculation of total energy, with supercell of 16, 32, 64 and 128 for NbMoTaW and 20, 40, 80 and 160 for NbMoTaWV and NbMoTaWTi respectively. Fig. \ref{fig:s3_0} shows the bcc supercell lattice of NbMoTaW (128 atoms), NbMoTaWV and NbMoTaWTi (160 atoms). For each supercell size, 200 configurations are randomly drawn and the corresponding energy are calculated by DFT method. Three smaller supercells with a total of 600 data are selected as the training dataset for Bayesian regularized regression and the largest supercell with 200 data are chosen for testing purpose. Six coordination shells in EPI model is chosen for this case. Fig. \ref{fig:s3_1} shows a comparison of predicted energy with DFT calculated energy for three HEAs and the corresponding training and testing RMSE are illustrated in Table \ref{tab:t1}. For these three HEAs, the testing RMSEs $\varepsilon_R \sim$ 0.6 meV show that the learned model is accurate and robust for a system described by a relatively large supercell size. 

\begin{table}[!ht] 
\centering
\caption{Training and testing RMSE accuracy of configurational energy for three HEAs}
\label{tab:t1}
\begin{tabular}{@{}cccc@{}}
\toprule
HEA   & \textup{NbMoTaW}  & \textup{NbMoTaWV}  & \textup{NbMoTaWTi} \\ \midrule
\textup{Training $\varepsilon_R$ (meV)} & 0.335 & 0.710 & 1.400 \\
\textup{Testing $\varepsilon_R$ (meV)} & 0.632 & 0.647 & 0.665 \\ \bottomrule
\end{tabular}
\end{table}

Typically, with the increasing of supercell size, the configurational systems tend to transit from ordered to disordered state. Due to the periodic boundary condition, the configurations drawn from smaller supercells often include different long-range order, while the samples obtained with larger supercells contain a various degree of short-range order. As a result, it is highly possible that a random system with large supercell is not well represented by only using samples generated from small supercells, which is commonly used because of its efficiency but may result in loss of physical information at the thermodynamic limit. To conduct a robust data-driven approach, we adopt an ensemble random sampling strategy that combines the data from different supercells. This simple yet efficient technique aims to obtain a training dataset with different degrees of order and disorder such that the data are more representative. 

\begin{figure}[!ht]   
  \centering
  {\includegraphics[width=0.5\textwidth]{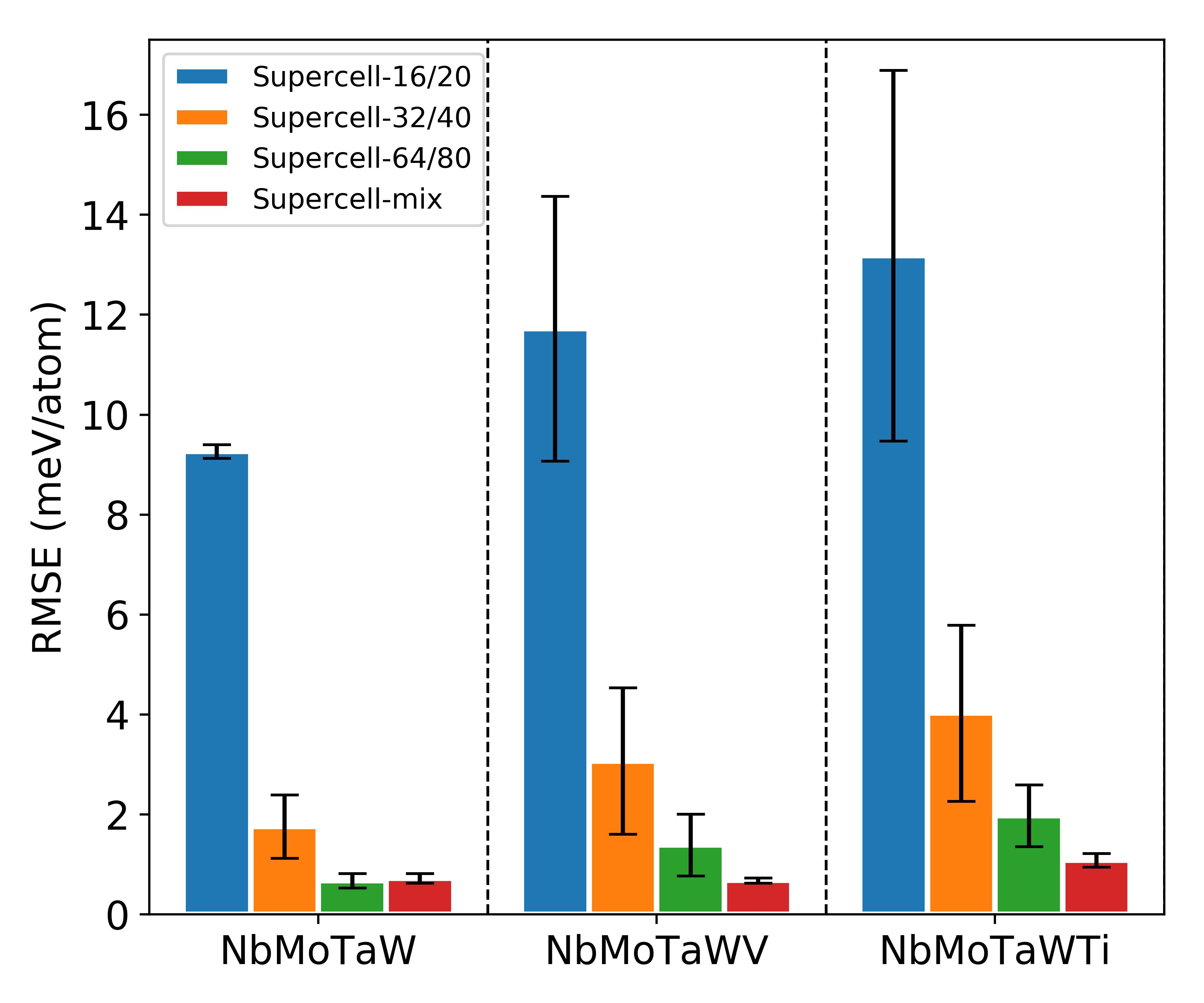}} \quad
  \caption{Testing performance comparison between ensemble sampling strategy and sampling drawn from only single supercell. Blue, orange and green bars: testing results using 150 data only from one specific supercell (16, 32, 64 for NbMoTaW and 20, 40, 80 for NbMoTaWV and NbMoTaWTi respectively). Red bar: testing results using an ensemble of 150 data that consists of three 50 data drawn from each supercell. } \label{fig:s3_2}
\end{figure} 

The benefit of ensemble sampling strategy can be seen in Fig. \ref{fig:s3_2}. All 200 data from relatively large supercell (128 for NbMoTaW and 160 for NbMoTaWV and NbMoTaWTi) are selected as the testing dataset. Total 150 data randomly drawn from each smaller supercells (16, 32, 64 for NbMoTaW and 20, 40, 80 for NbMoTaWV and NbMoTaWTi) are selected as the training dataset. The ensemble sampling strategy performs an ensemble of dataset which is achieved by randomly drawn 50 data using Latin hypercube sampling \cite{shields2016generalization} from each of three supercells. In terms of these four cases, 100 random trials are carried out to estimate the standard deviation (error bar in Fig. \ref{fig:s3_2}) of the testing RMSE results. It can be easily seen that the mean and standard deviation of RMSE results are significantly large when only using the smallest supercell. The results underscore a fact that the configuration space in a random system is not well covered by training data only drawn from small supercells. The performance of 32 (40) and 64 (80) are better than that of 16 (20)-atom supercell, showing that more degrees of short-range and long-range order are captured by these training data. The ensemble sampling strategy (red bars) showing a minimal RMSE mean ($<$1 meV) and standard deviation beat the other three cases in terms of the accuracy and stability. This robust strategy plays a substantial role in the data-driven modeling of the configurational energy such that the subsequent Monte Carlo simulation based on this efficient Hamiltonian can safely explore the whole regions of configuration space. 

\subsection{Uncertainty in effective pair interaction bonds}

\begin{figure}[!ht]   
  \centering
  {\includegraphics[width=0.8\textwidth]{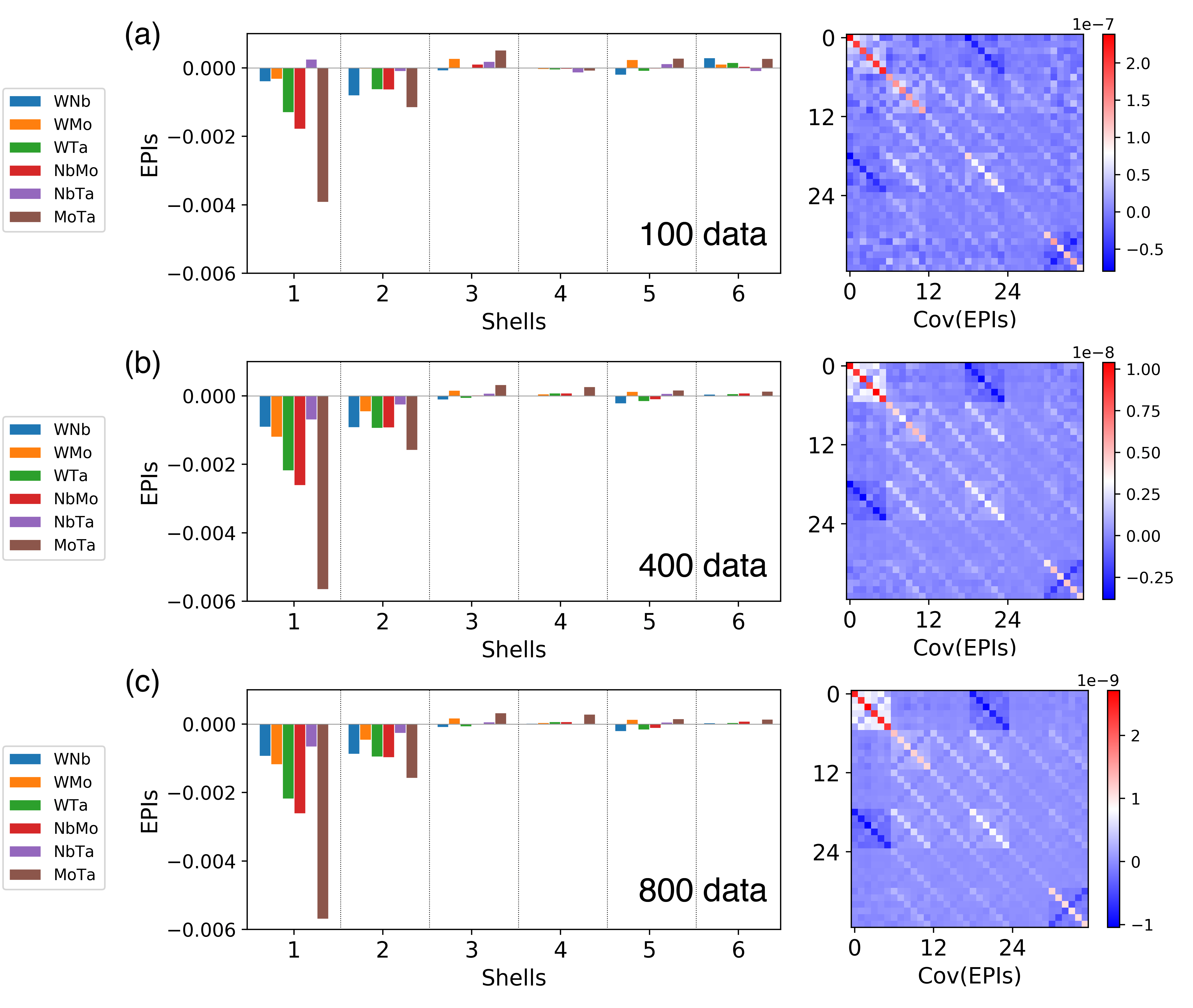}} \quad
  \caption{Effective pair interaction (EPI) bonds and their uncertainties that are quantified by the variance-covariance matrix given different sizes of training dataset for NbMoTaW} \label{fig:s3_3}
\end{figure}

\begin{figure}[!ht]   
  \centering
  {\includegraphics[width=0.8\textwidth]{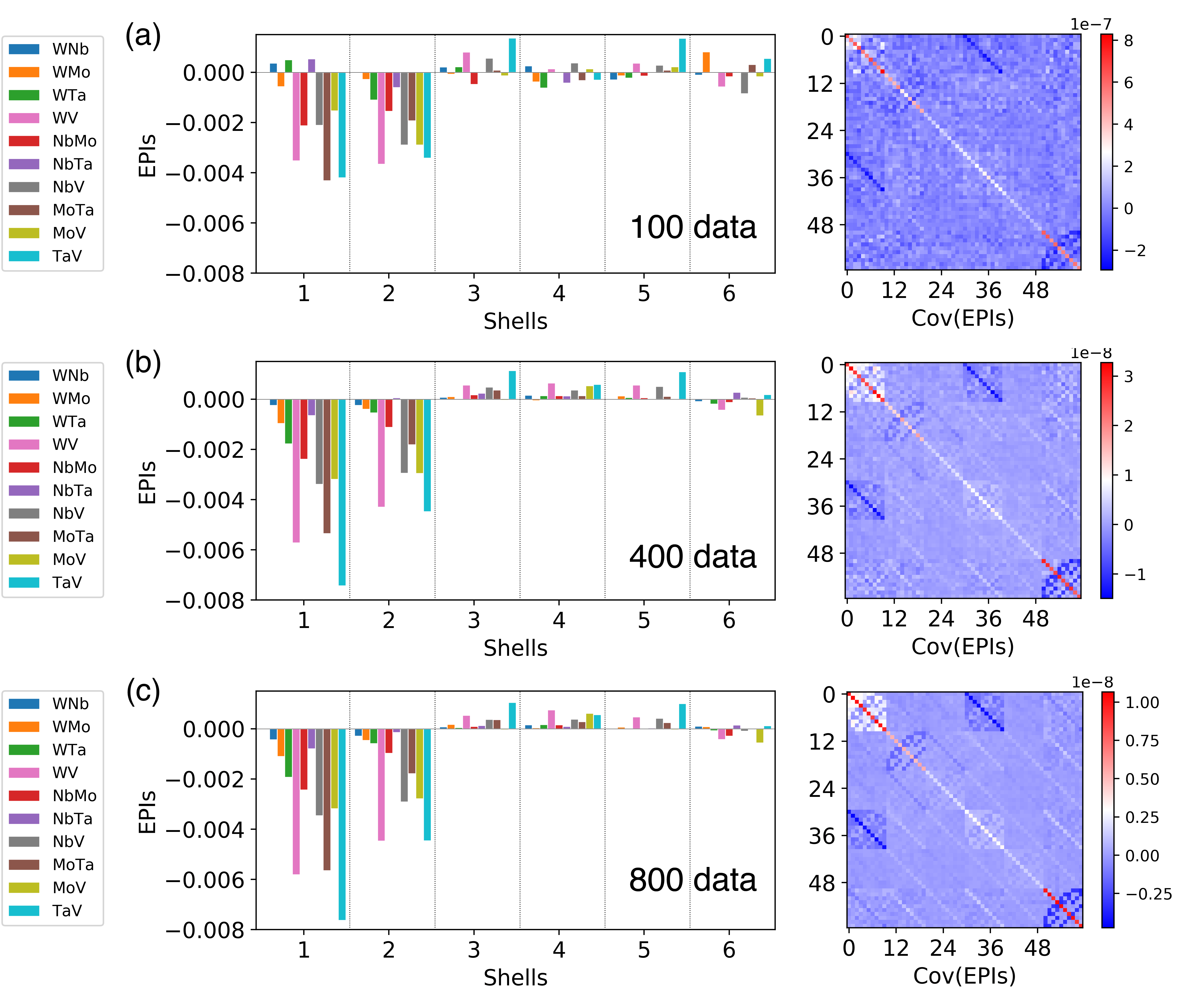}} \quad
  \caption{Effective pair interaction (EPI) bonds and their uncertainties that are quantified by the variance-covariance matrix given different sizes of training dataset for NbMoTaWV} \label{fig:s3_4}
\end{figure}

\begin{figure}[!ht]   
  \centering
  {\includegraphics[width=0.8\textwidth]{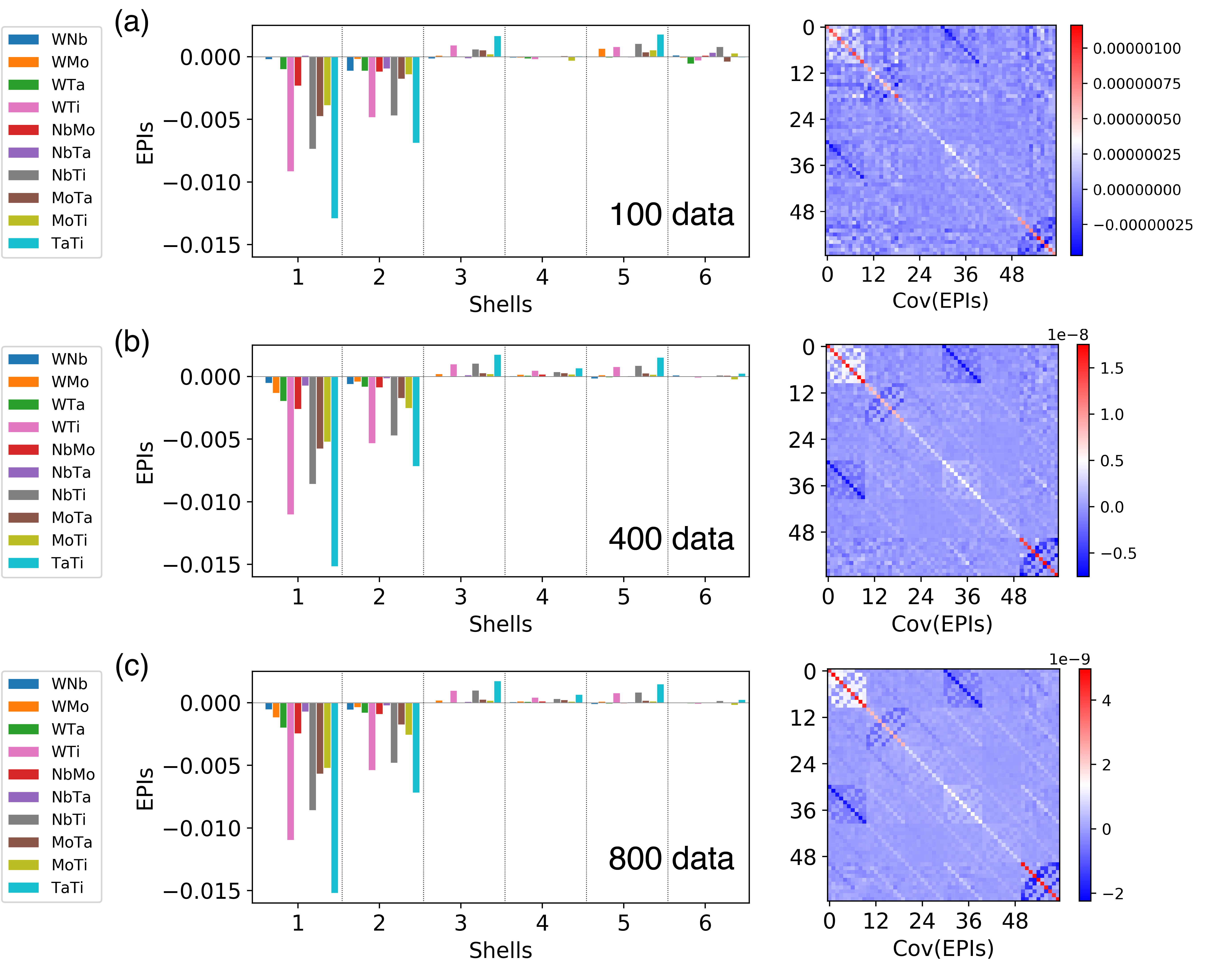}} \quad
  \caption{Effective pair interaction (EPI) bonds and their uncertainties that are quantified by the variance-covariance matrix given different sizes of training dataset for NbMoTaTi} \label{fig:s3_5}
\end{figure}

The coordination shells, as the physical features, have a pivotal influence on the EPI model and their impact can be analyzed from the EPI parameters, as shown in Fig. \ref{fig:s3_3} - Fig. \ref{fig:s3_5}. For all three refractory HEAs, the first two shells, involving the nearest and next-nearest neighbor interactions are dominant, while the 3rd to 6th shells present a less essential role. A comparison between NbMoTaW and the other two HEAs, NbMoTaWV and NbMoTaWTi shows that the EPIs of NbMoTaW, as shown in Fig. \ref{fig:s3_3} (c), is relatively stable and short-ranged interacted due to small magnitude associated with the long-ranged shells, while the NbMoTaWV and NbMoTaWTi, as shown in Fig. \ref{fig:s3_4} (c) and Fig. \ref{fig:s3_5} (c) are more frustrated and long-ranged, with certain contribution from up to the 6th shell. 

Moreover, three different dataset sizes, $n_d=100, 400$ and 800 are shown here to investigate the effect of data on the EPI parameters (chemical bonds). The ensemble sampling strategy is used herein such that we collect 25, 100 and 200 data from each of four supercell sizes, i.e. 16, 32, 64 and 128 for NbMoTaW and 20, 40, 80 and 160 for NbMoTaWV and NbMoTaWTi. Given small dataset size, for example, $n_d=100$, the trend of EPI bonds is consistent - the first two shells are dominant, but the values still have a discrepancy from larger dataset size, for example, $n_d=800$. In other words, the uncertainties associated with the EPI bonds are primarily caused by a lack of data. This is reflected in the variance-covariance matrix of EPI bonds, as shown in Fig. \ref{fig:s3_3} - Fig. \ref{fig:s3_5} (a). It is easily seen that the variance in 1st shells is the largest, followed by the 2nd shell and the 6th shell, which are larger than the other shells. The covariance values trending to zero demonstrates that there is no strong correlation among each of shells, which is agreed with the independent assumption. 

As the data set size increases, from $n_d=100$ to $n_d=400$, the EPI bonds become more stable and almost same as the case of $n_d=800$. The corresponding variance is also significantly reduced by more than one order of magnitude, as shown in Fig. \ref{fig:s3_3} - Fig. \ref{fig:s3_5} (b). Furthermore, the variance-covariance matrix presents an increasingly clear ``pattern" with the increasing of dataset size and we therefore have some observations: a) it is clear to see that the pattern of variance-covariance matrix is divided by the identified physical feature (coordination shell); b) the EPI bonds in the nearest-neighbor shell shows a stronger positive correlation, while the 6th shell illustrates a relatively large uncertainty with a negative correlation among the EPI bonds; c) the EPI bonds between the 1st shell and 4th shell have a negative correlation. Even though most of the EPI bonds are physically independent as we expected, the EPI bonds in a specific shell are not completely independent, particularly the nearest-neighbor shell, and there is still slightly either positive or negative correlation between different shells. 

\subsection{Effect of physical feature selection}

\begin{figure}[!ht]   
  \centering
  \subfigure[]{\includegraphics[width=0.3\textwidth]{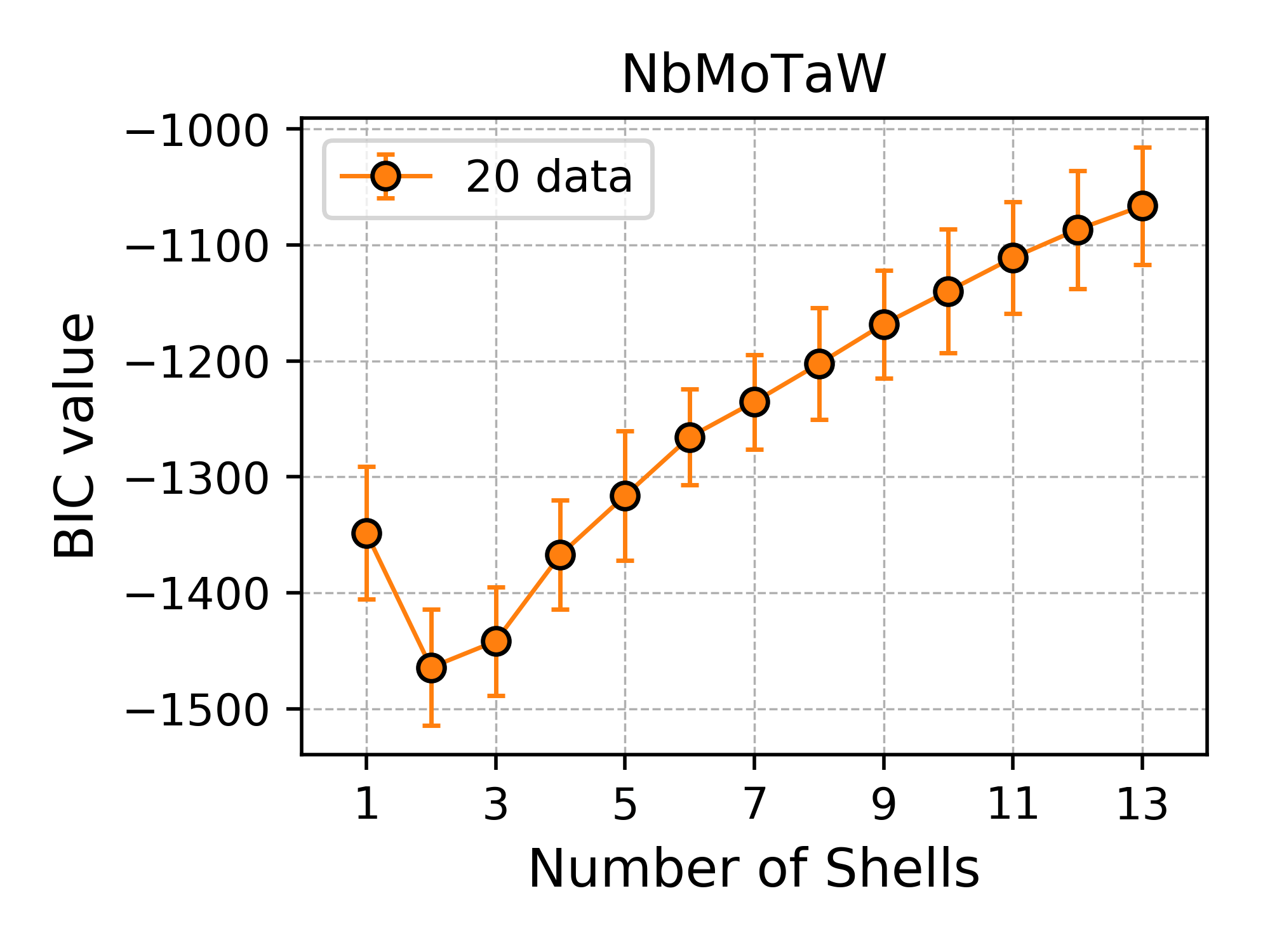}}
  \subfigure[]{\includegraphics[width=0.3\textwidth]{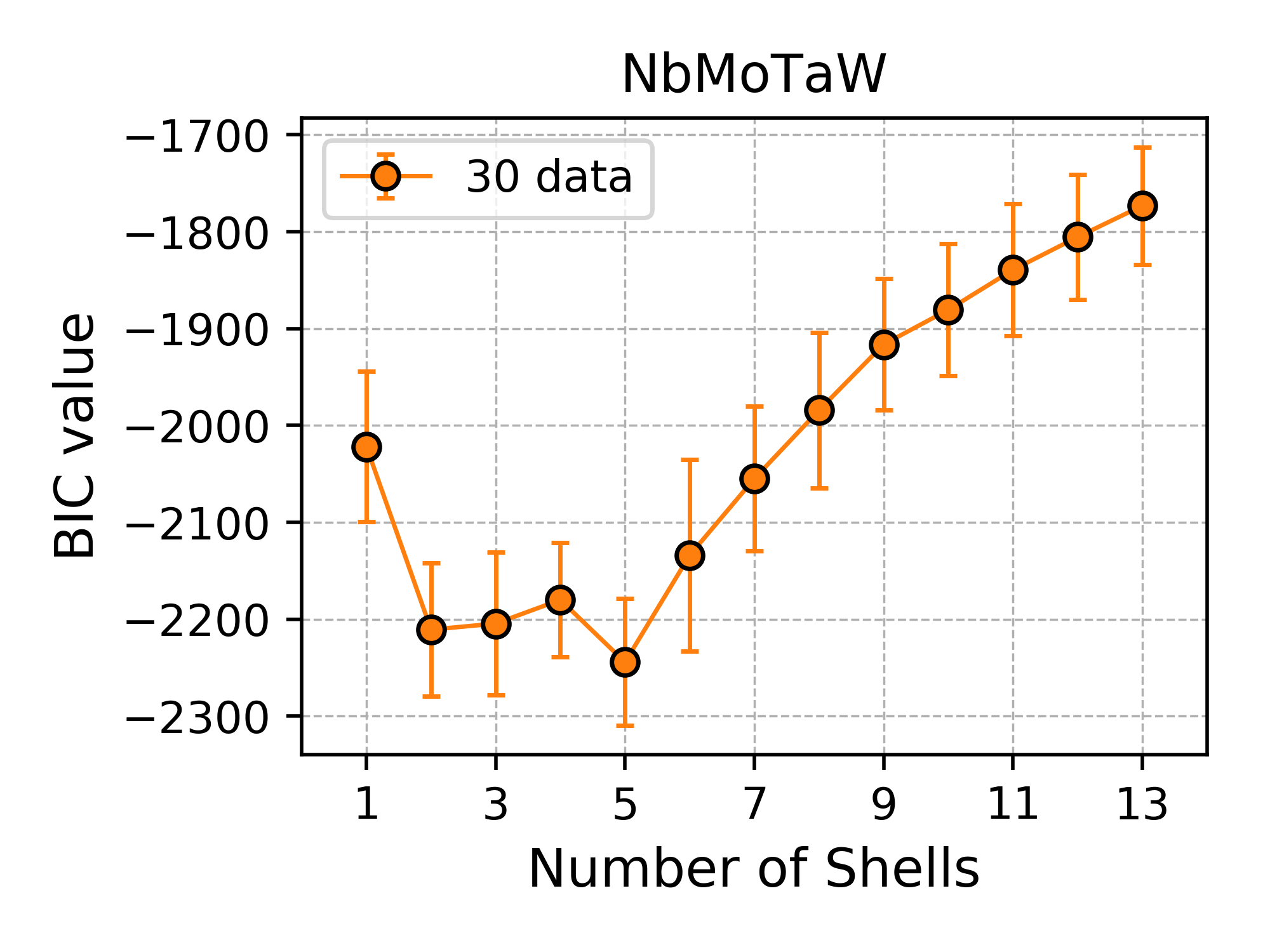}}
  \subfigure[]{\includegraphics[width=0.3\textwidth]{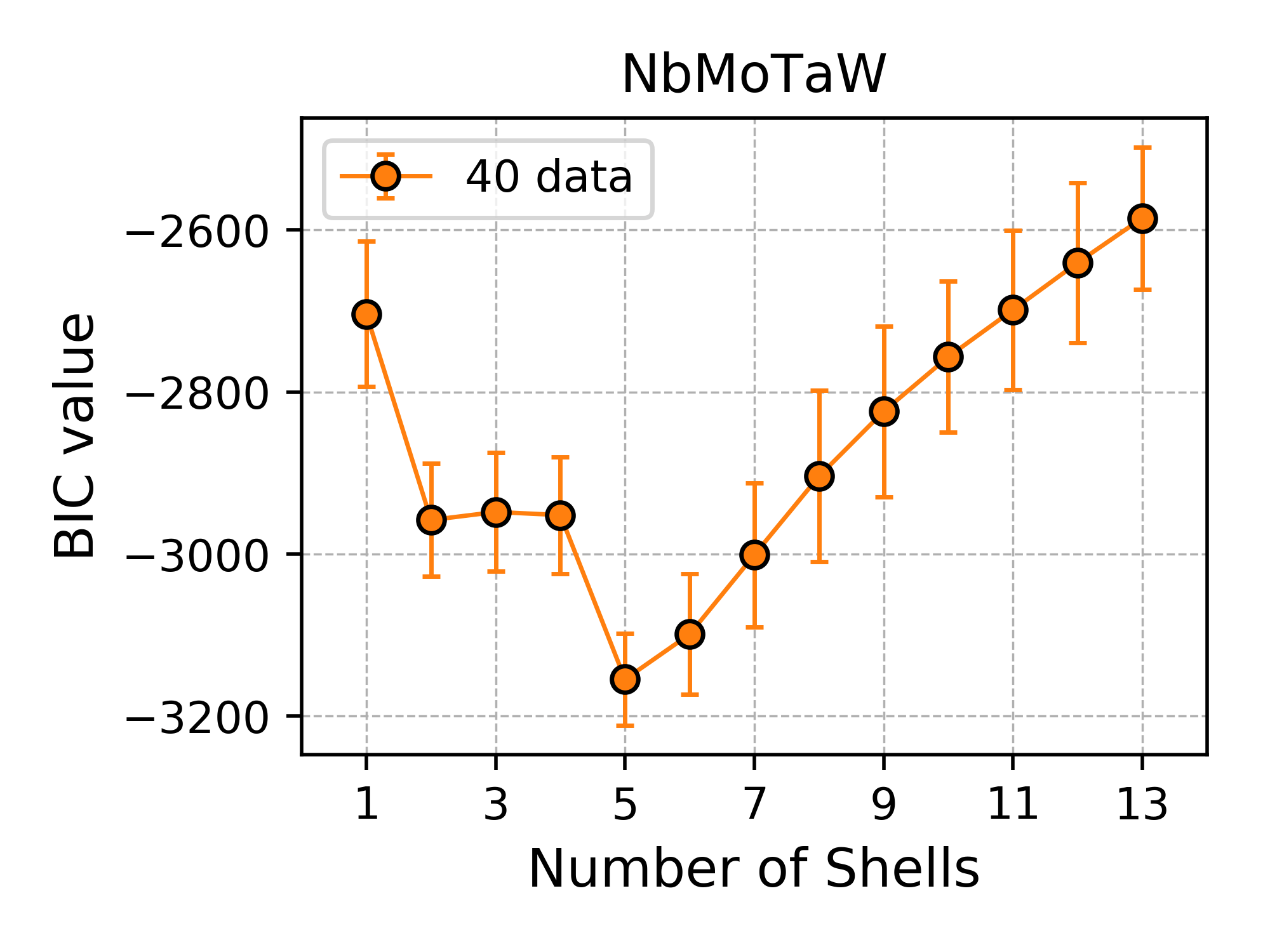}} \\
  \subfigure[]{\includegraphics[width=0.3\textwidth]{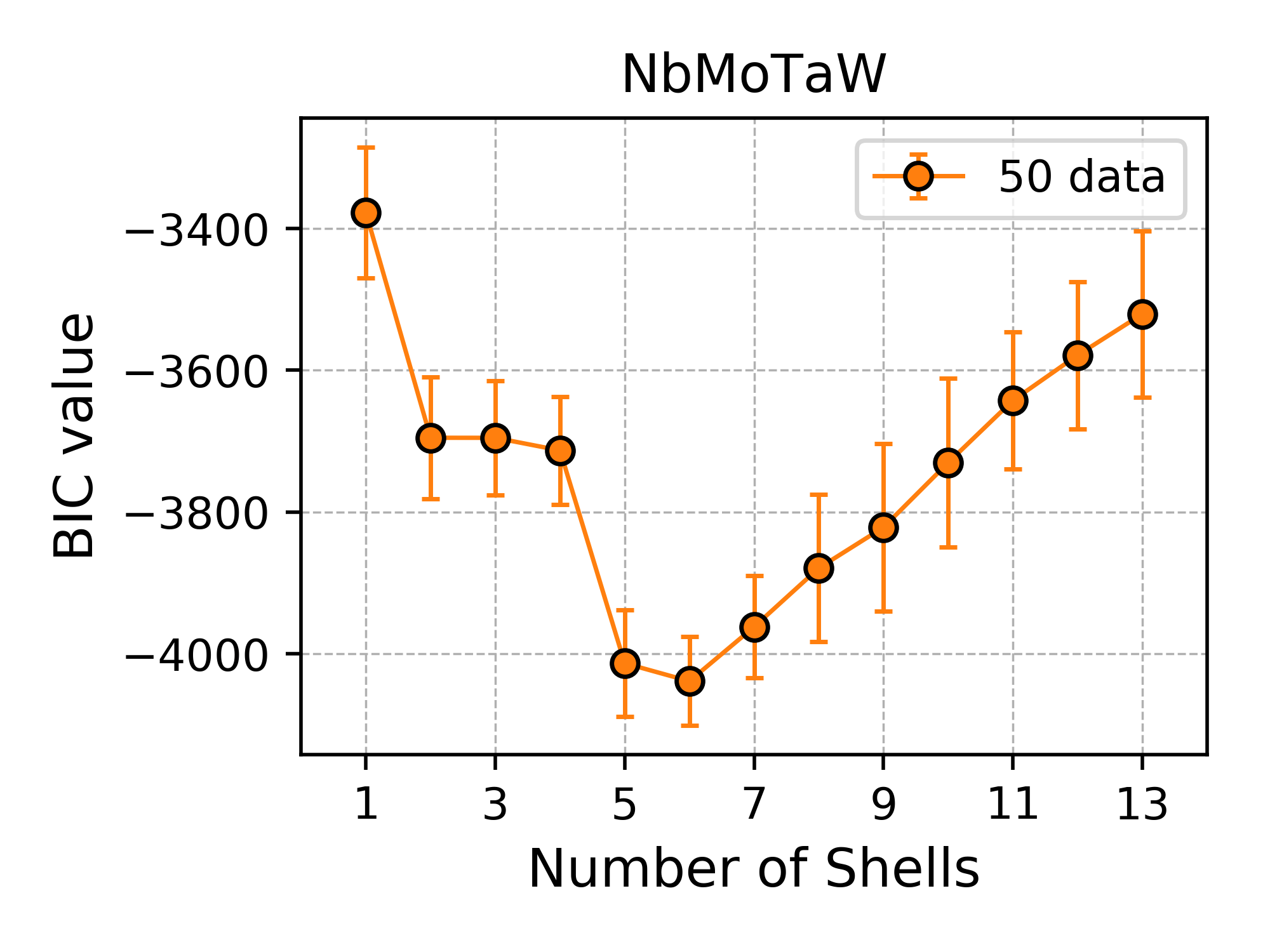}}
  \subfigure[]{\includegraphics[width=0.3\textwidth]{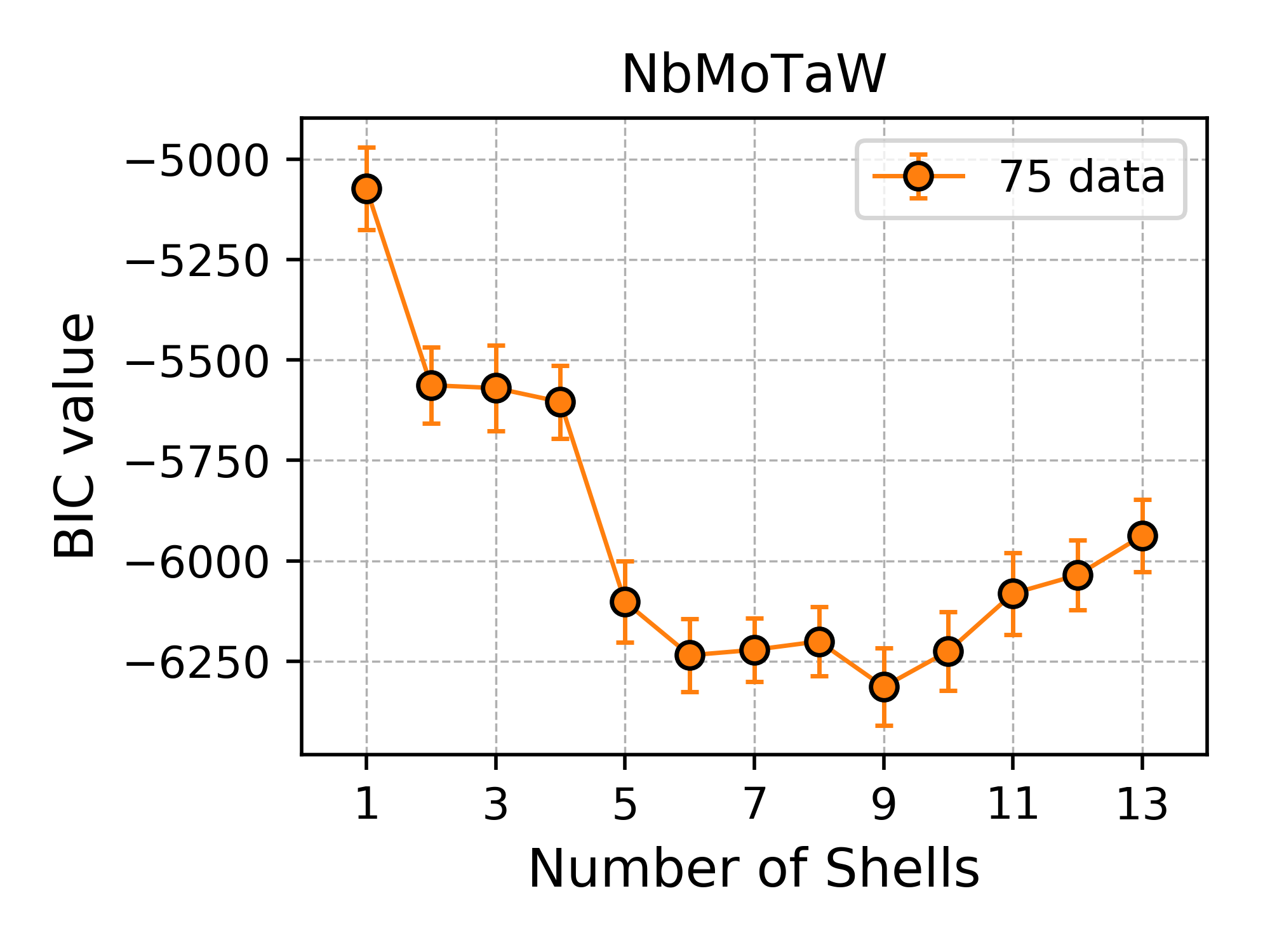}}
  \subfigure[]{\includegraphics[width=0.3\textwidth]{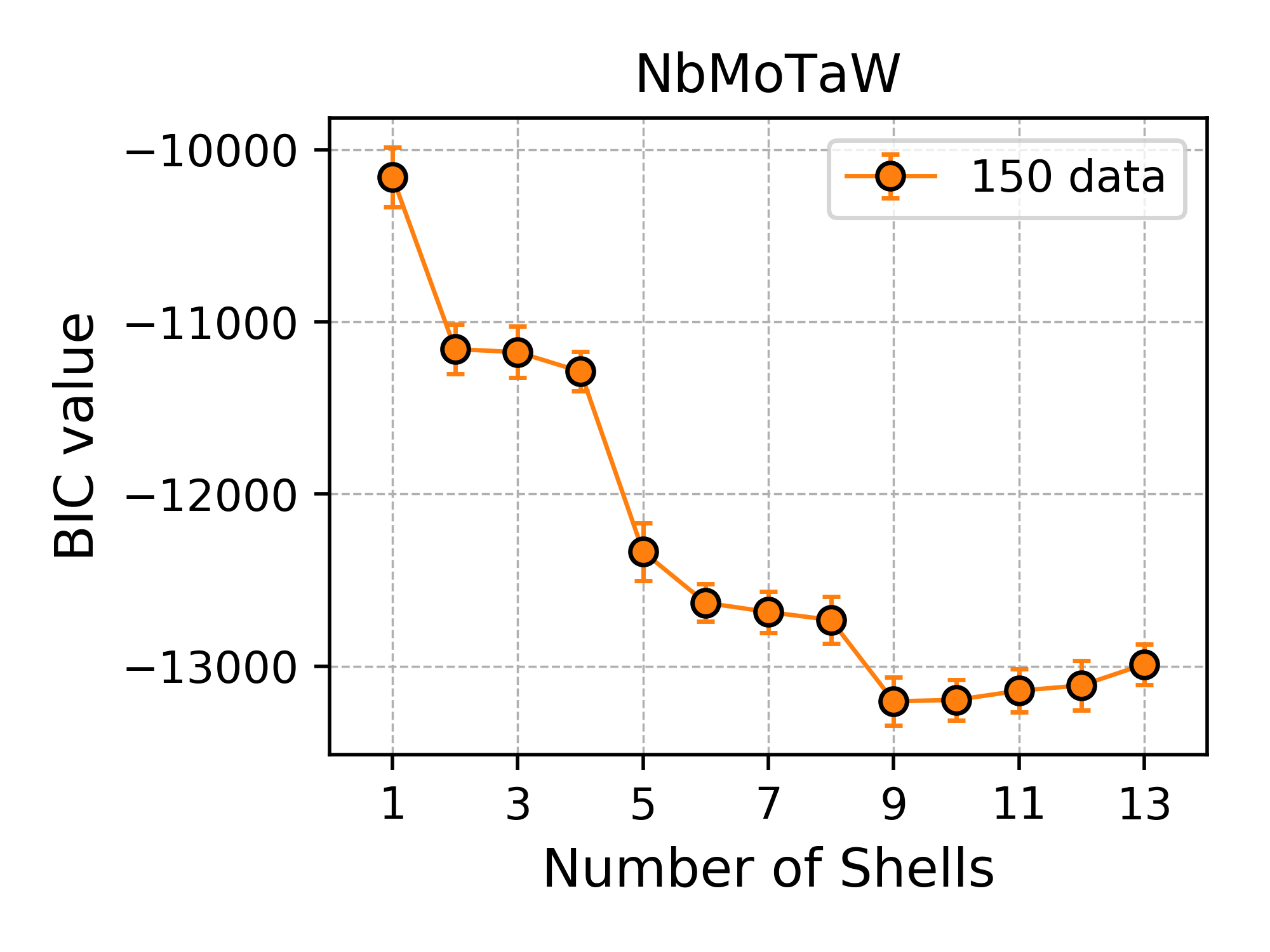}}
  \caption{Bayesian information criterion results for different number of shells in NbMoTaW given a specific training dataset size, including (a) 20 data, (b) 30 data, (c) 40 data, (d) 50 data, (e) 75 data and (f) 150 data} \label{fig:s33_1}
\end{figure}

\begin{figure}[!ht]   
  \centering
  \subfigure[]{\includegraphics[width=0.3\textwidth]{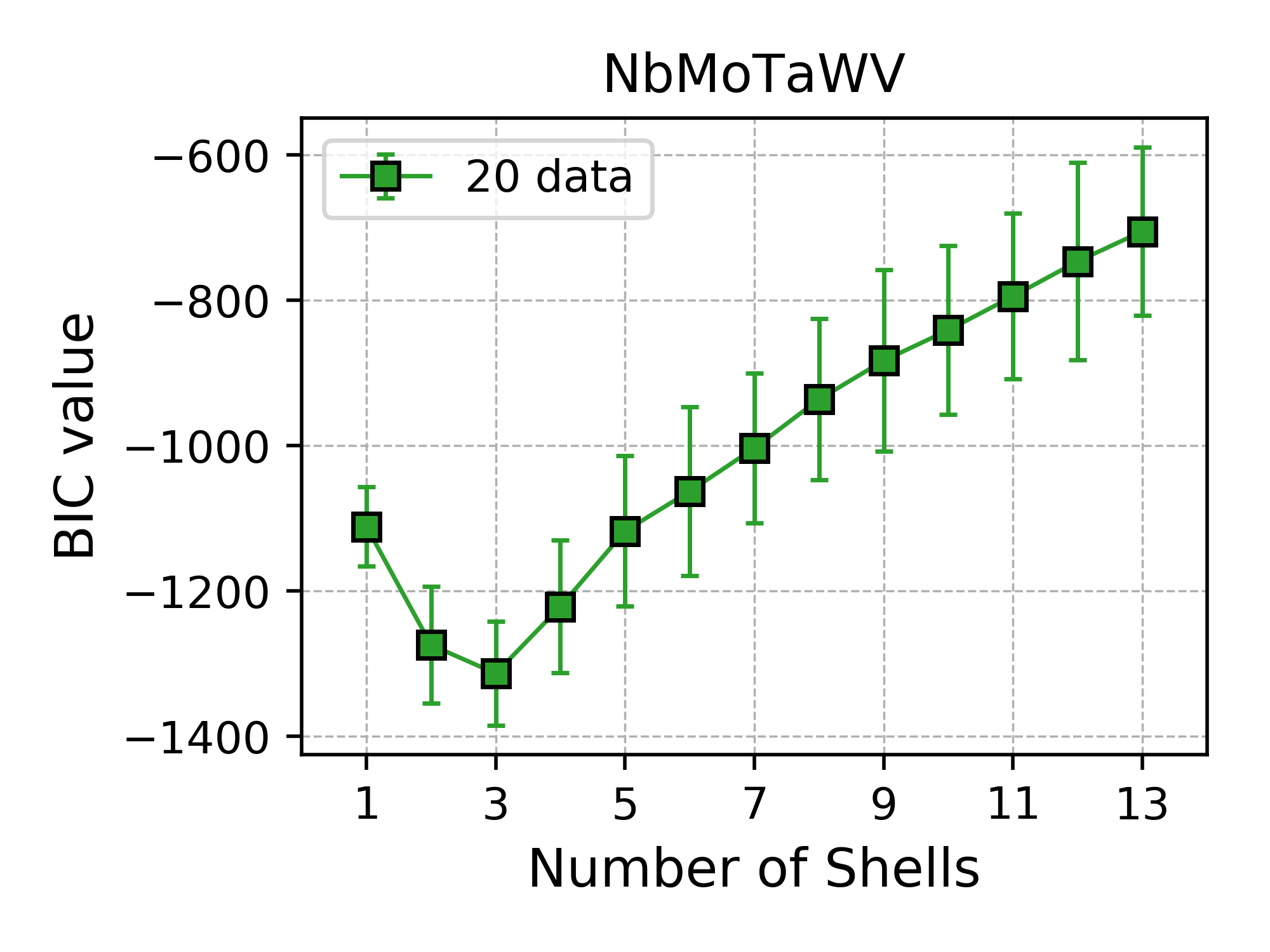}}
  \subfigure[]{\includegraphics[width=0.3\textwidth]{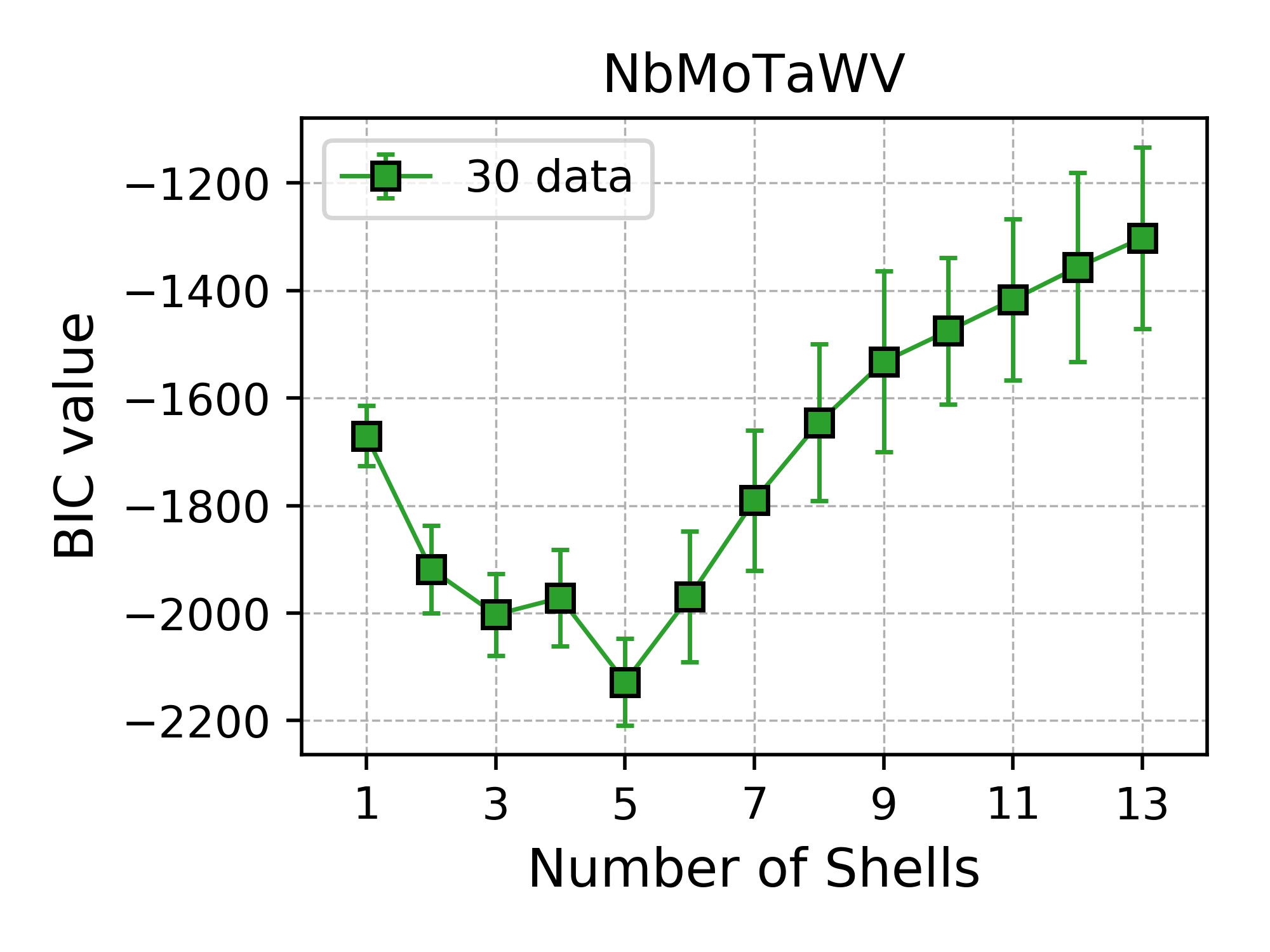}}
  \subfigure[]{\includegraphics[width=0.3\textwidth]{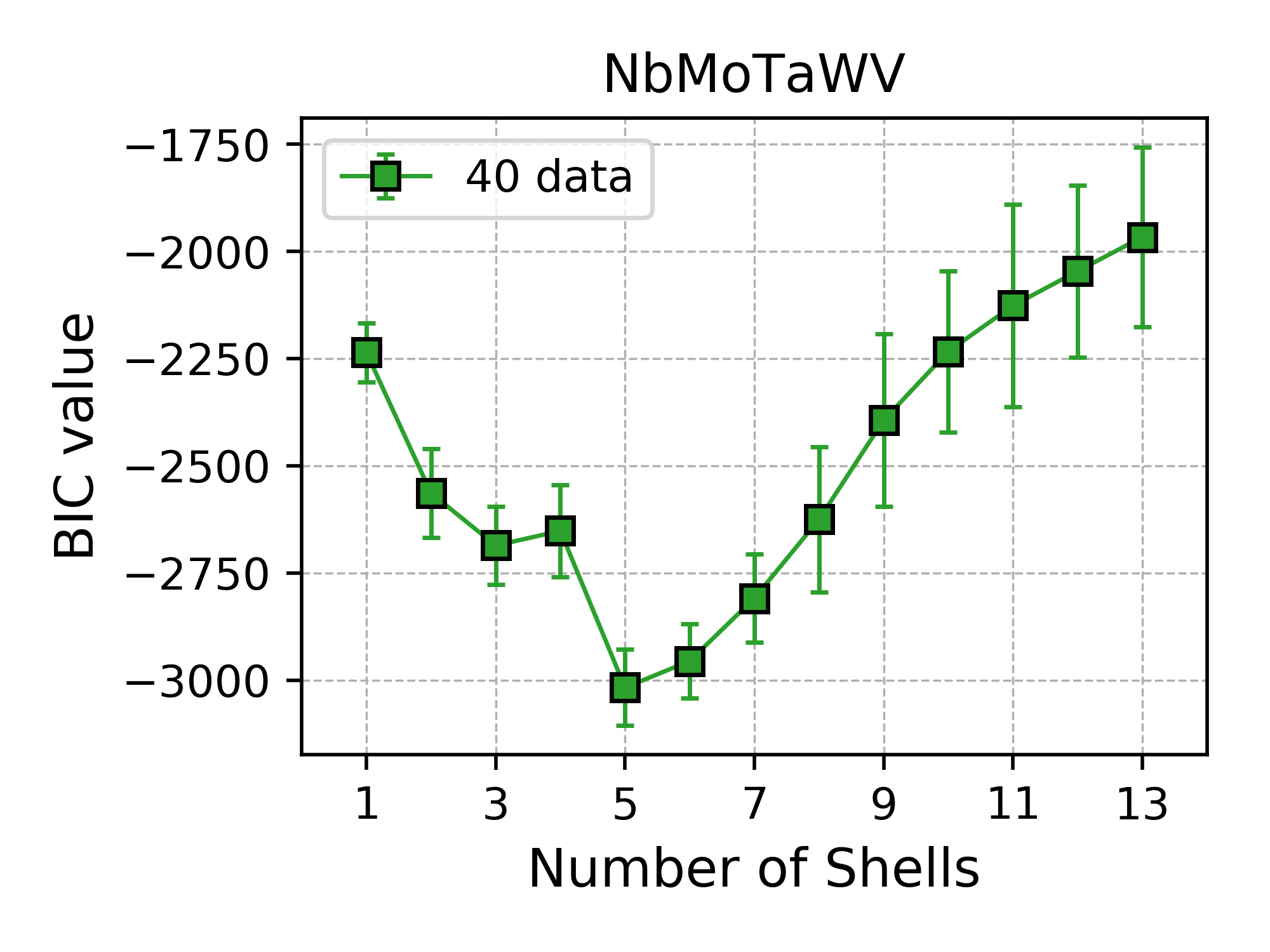}} \\
  \subfigure[]{\includegraphics[width=0.3\textwidth]{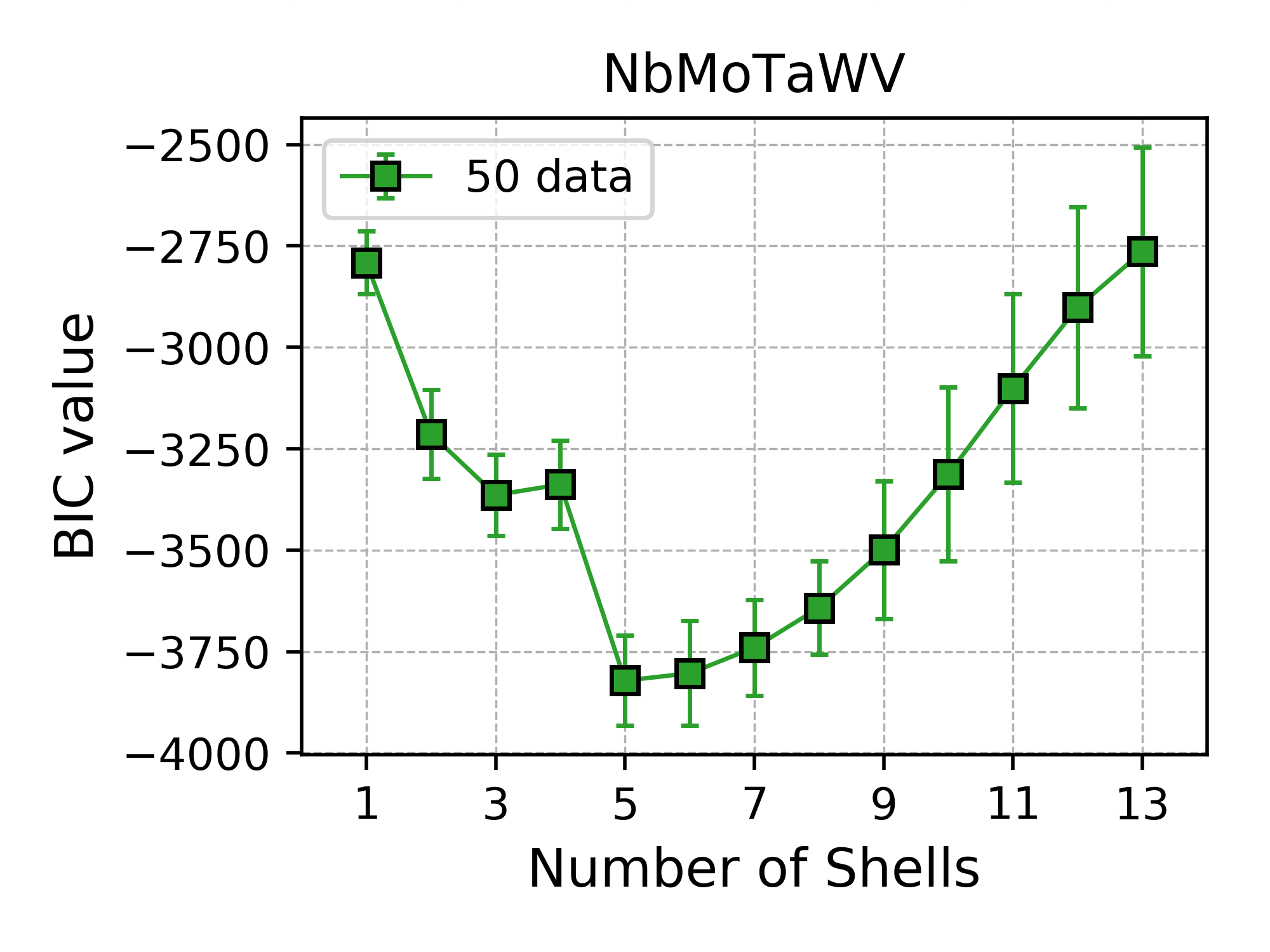}}
  \subfigure[]{\includegraphics[width=0.3\textwidth]{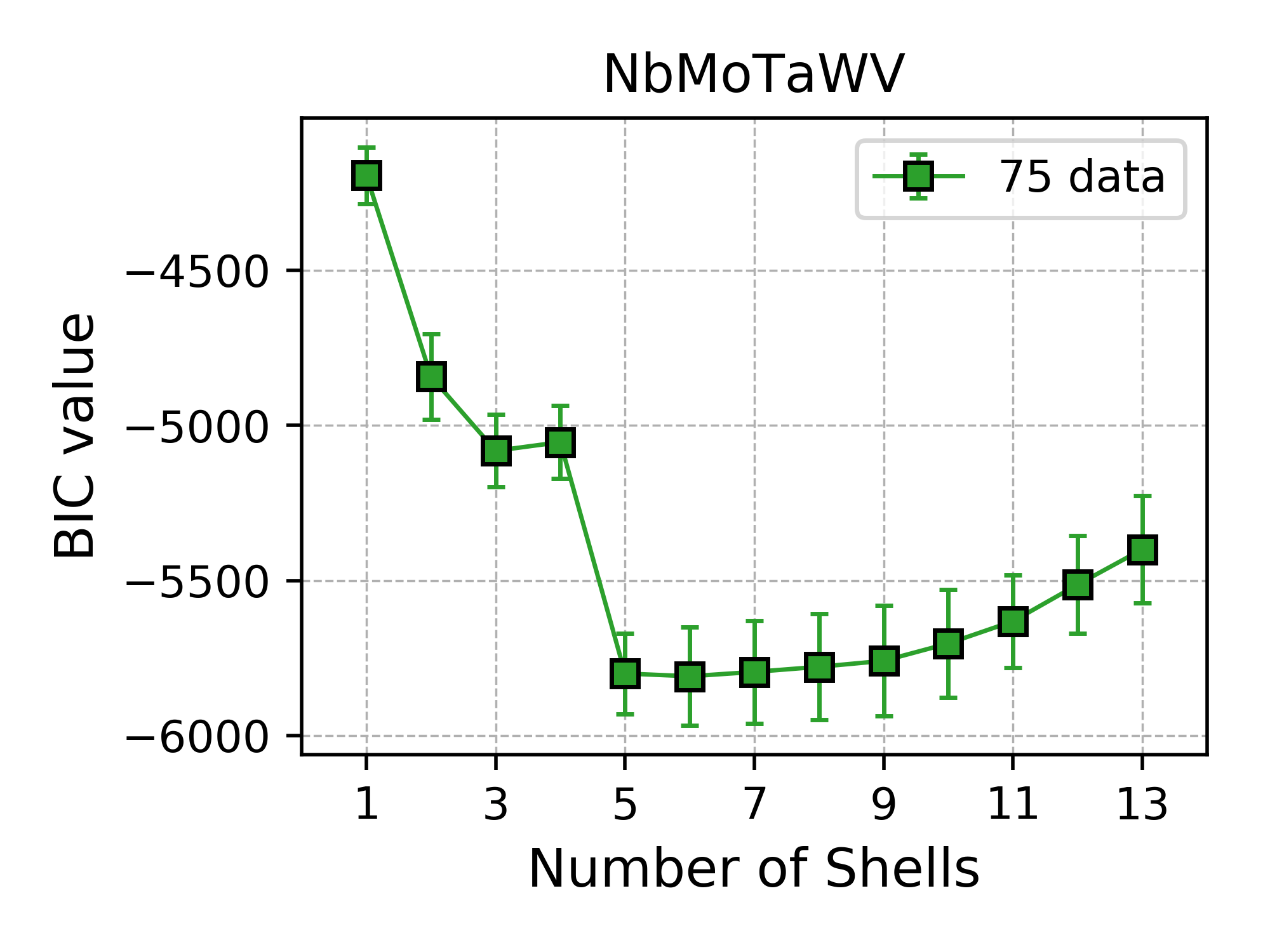}}
  \subfigure[]{\includegraphics[width=0.3\textwidth]{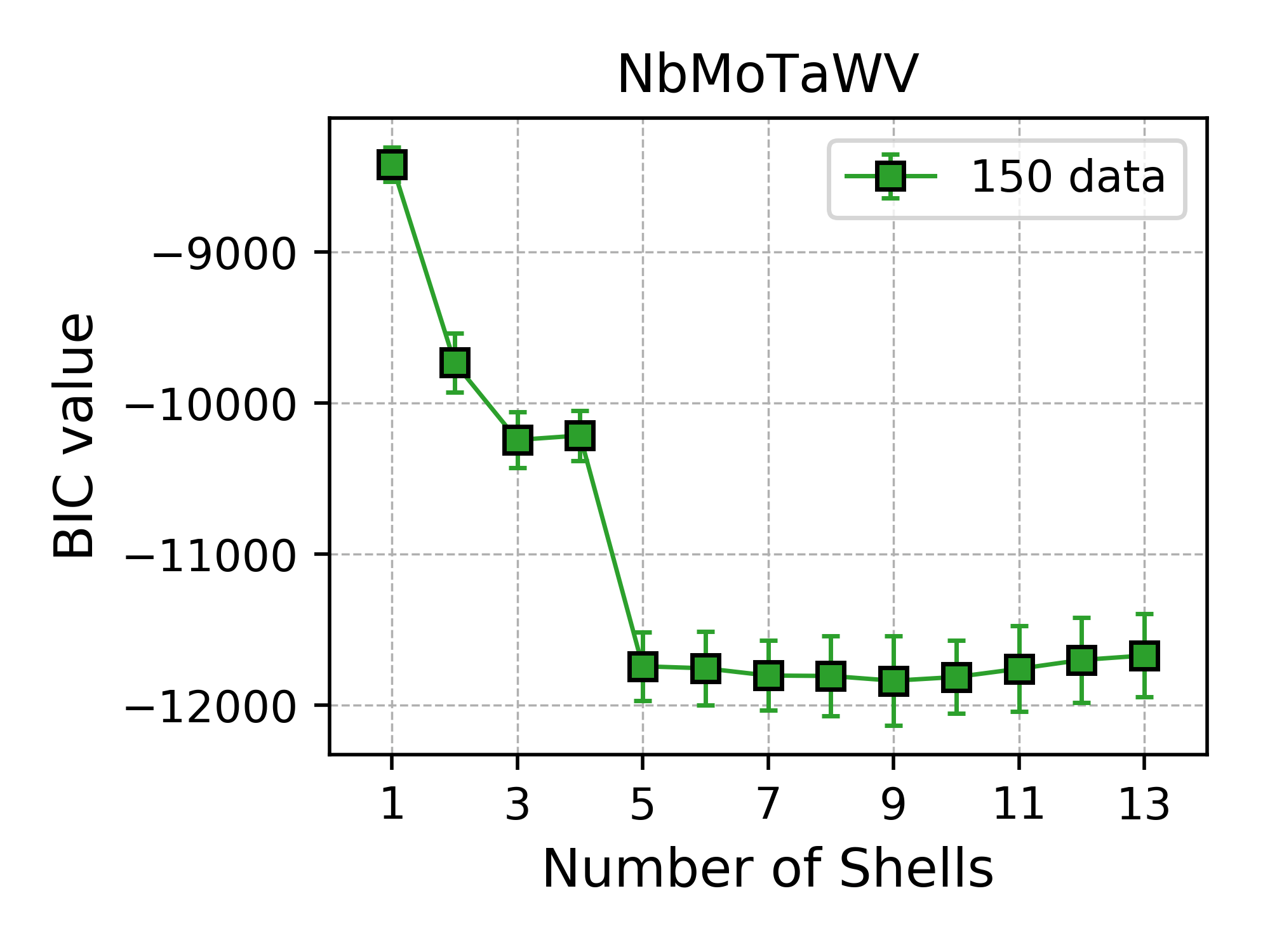}}
  \caption{Bayesian information criterion results for different number of shells in NbMoTaWV given a specific training dataset size, including (a) 20 data, (b) 30 data, (c) 40 data, (d) 50 data, (e) 75 data and (f) 150 data} \label{fig:s33_2}
\end{figure}

\begin{figure}[!ht]   
  \centering
  \subfigure[]{\includegraphics[width=0.3\textwidth]{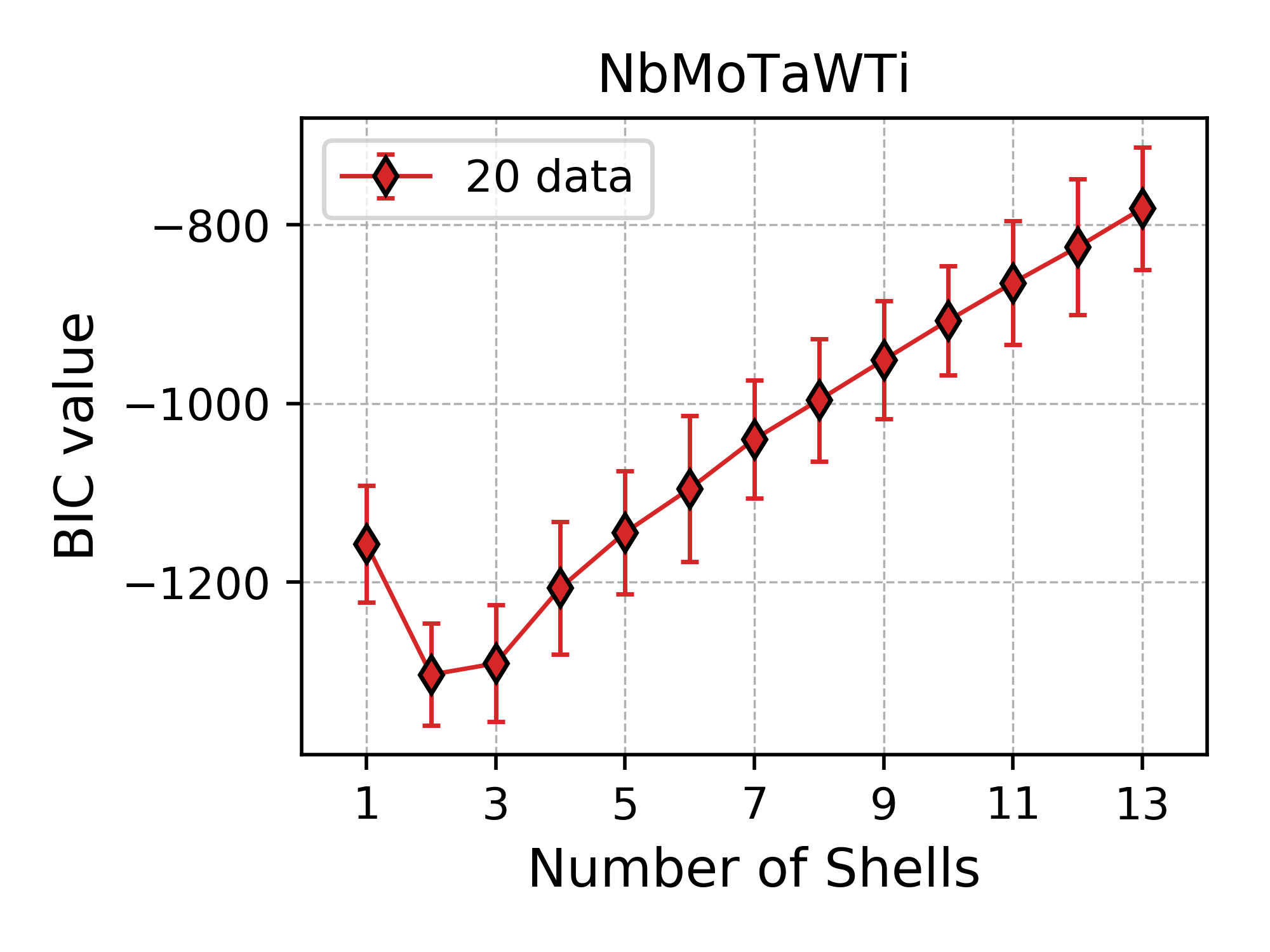}}
  \subfigure[]{\includegraphics[width=0.3\textwidth]{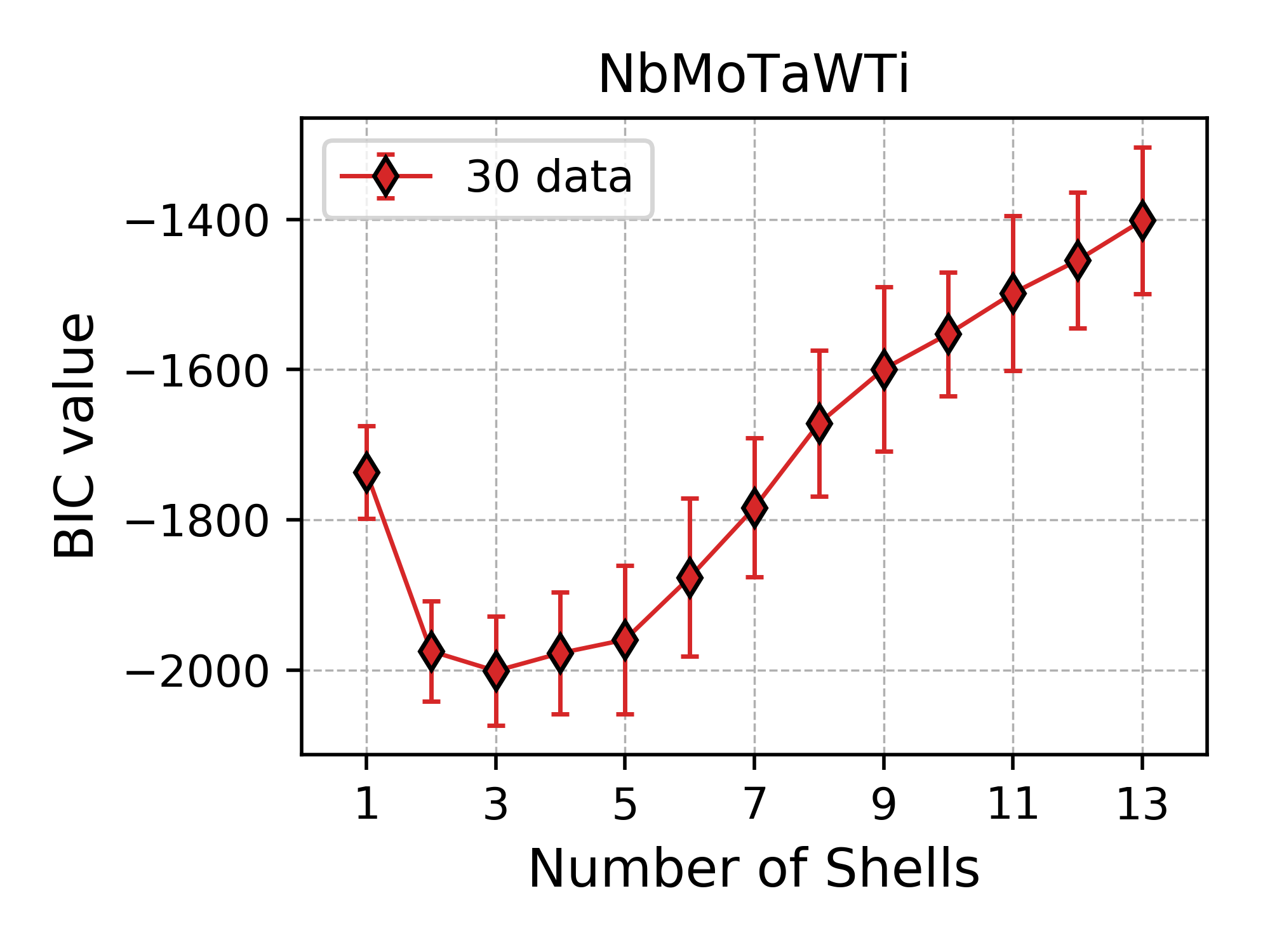}}
  \subfigure[]{\includegraphics[width=0.3\textwidth]{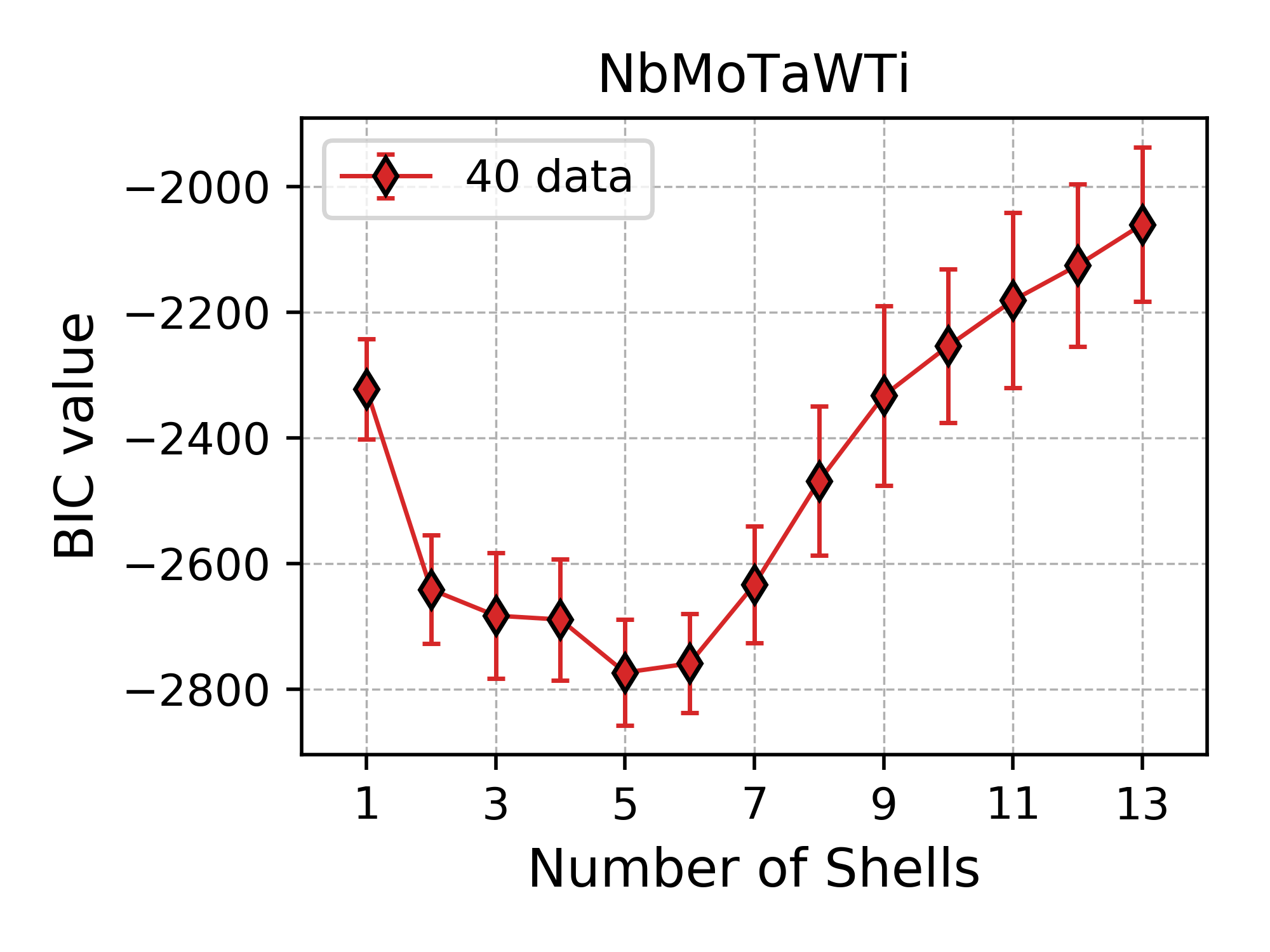}} \\
  \subfigure[]{\includegraphics[width=0.3\textwidth]{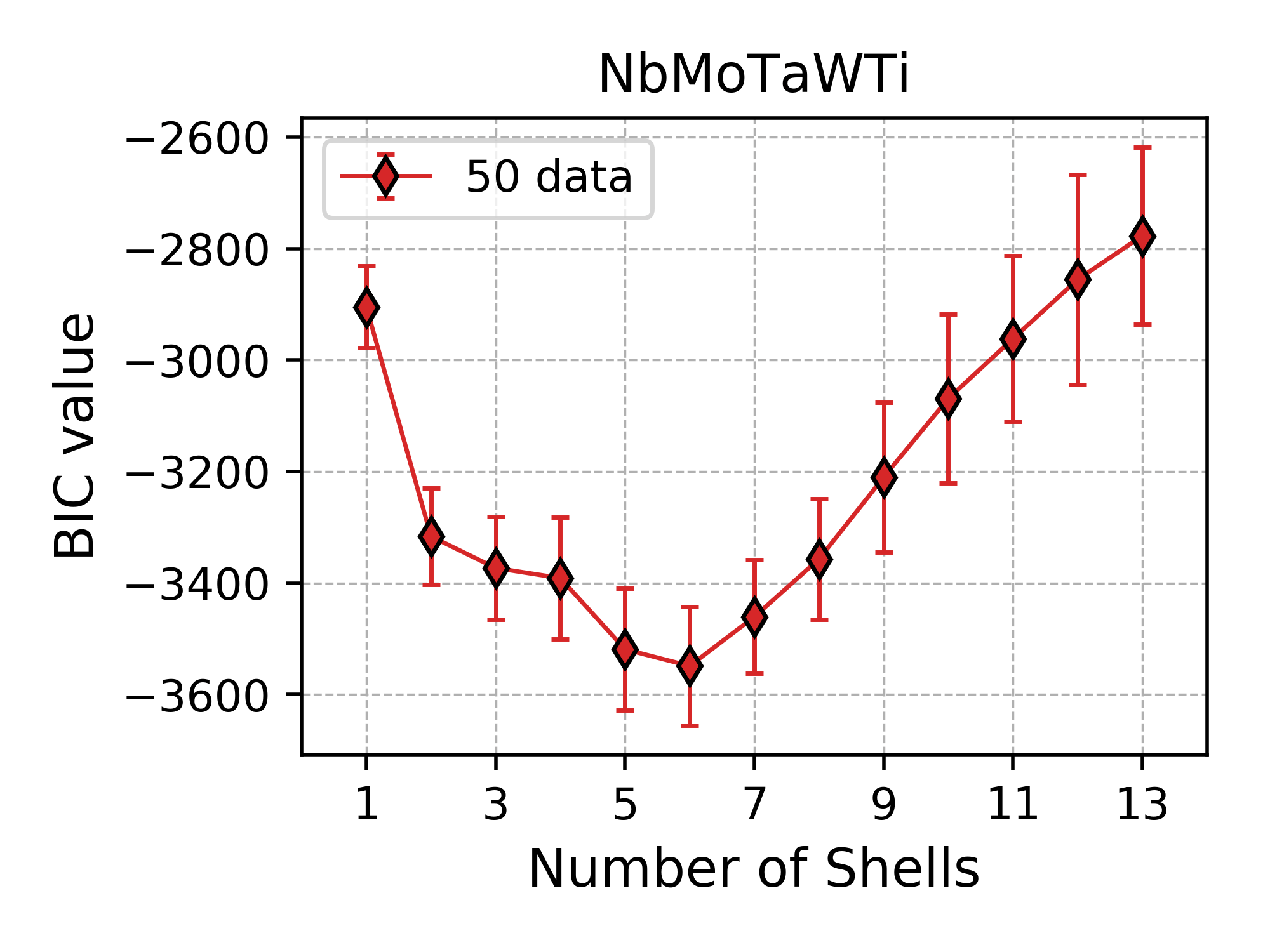}}
  \subfigure[]{\includegraphics[width=0.3\textwidth]{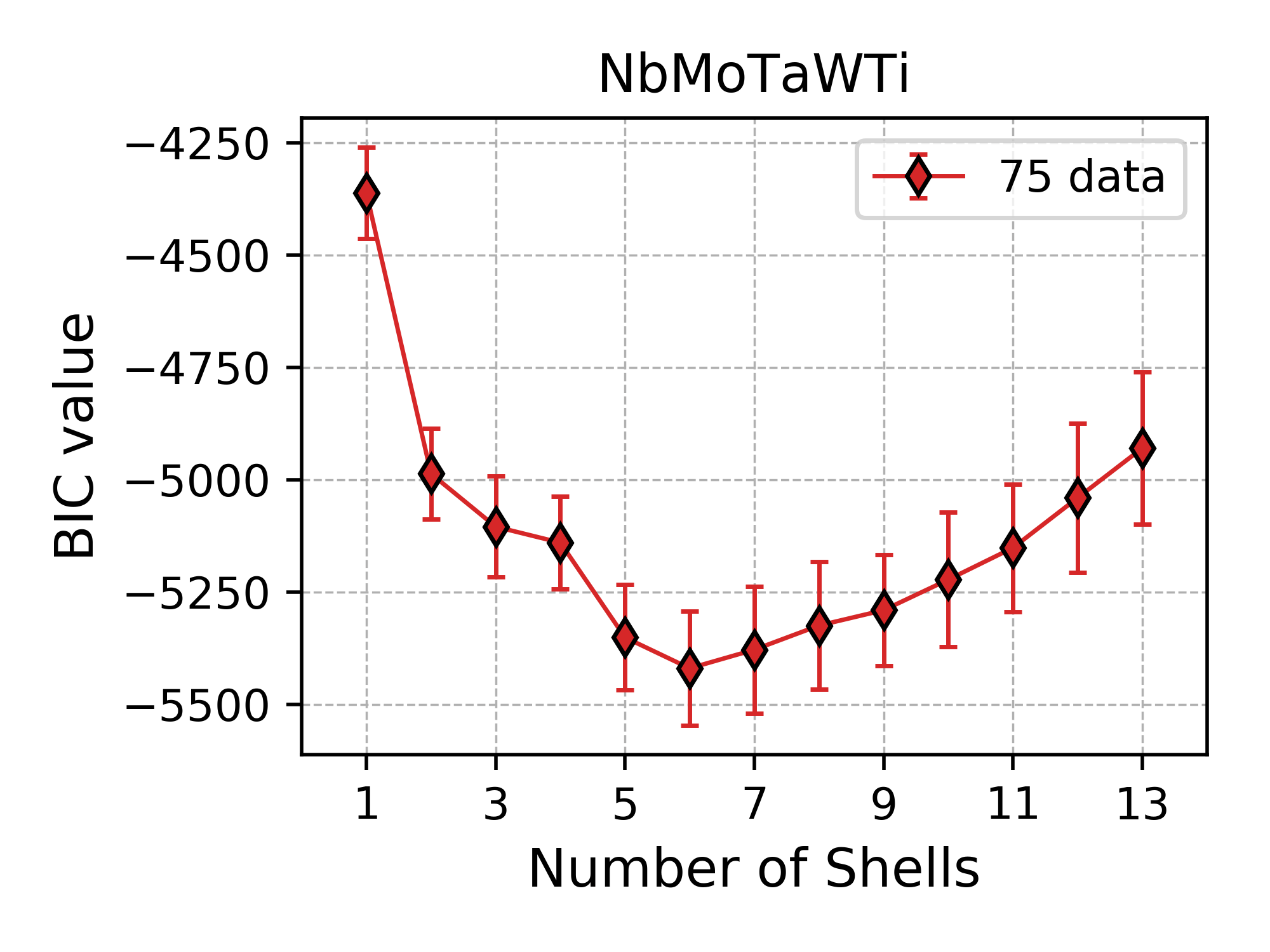}}
  \subfigure[]{\includegraphics[width=0.3\textwidth]{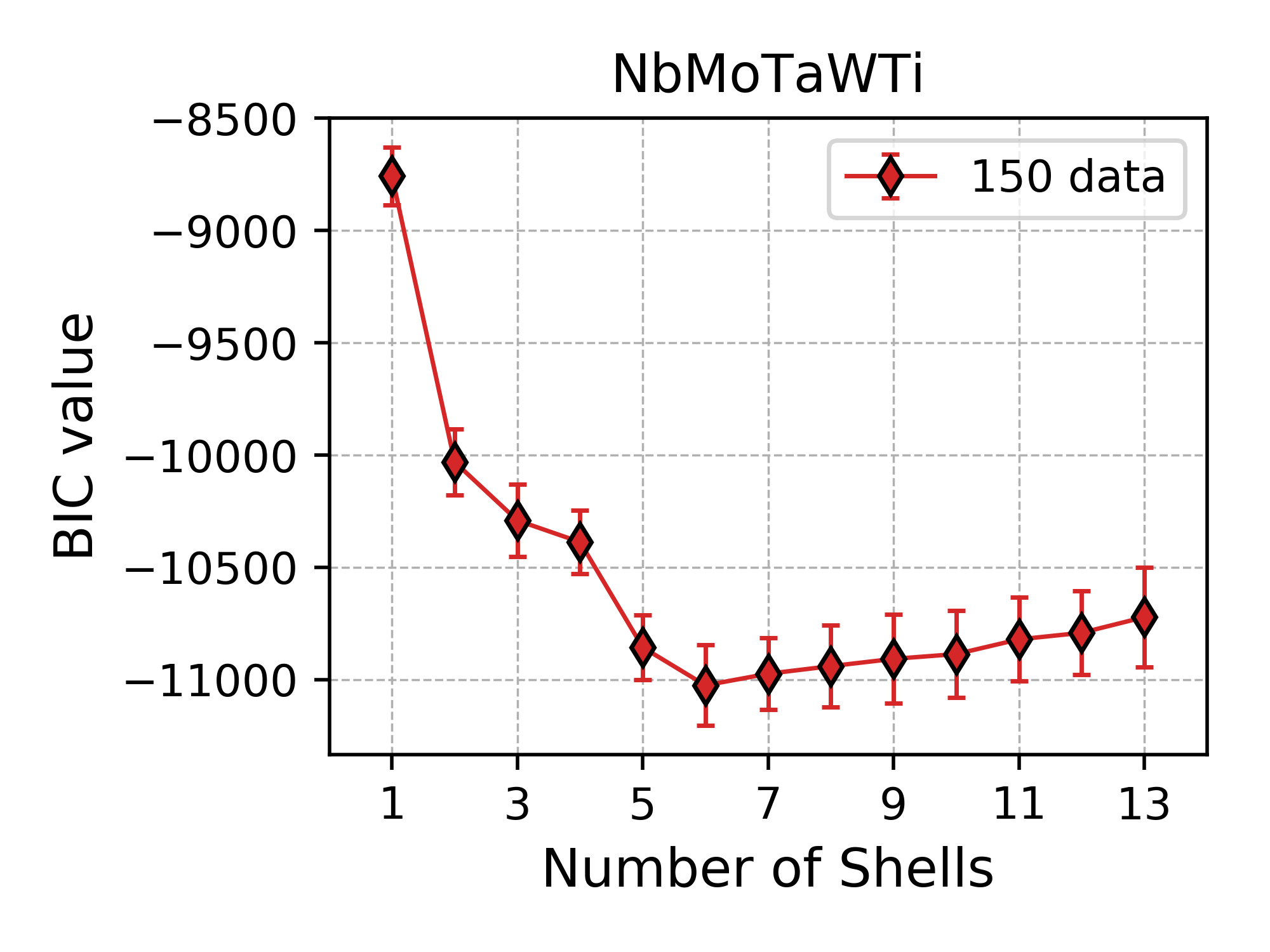}}
  \caption{Bayesian information criterion results for different number of shells in NbMoTaWTi given a specific training dataset size, including (a) 20 data, (b) 30 data, (c) 40 data, (d) 50 data, (e) 75 data and (f) 150 data} \label{fig:s33_3}
\end{figure}

Due to lack of data, the uncertainties of the EPI bonds have been clearly quantified via Bayesian regularized regression as discussed above and these uncertainties will be propagated to the prediction of configurational energy and finally affect the thermodynamic quantities. It therefore gives rise to a critical issue that is how to systemically reduce the uncertainty and improve the robustness and reliability of prediction when only limited data are collected. To deal with this challenge, we introduce Bayesian information criterion (BIC) to conduct a physical feature selection and investigate the effect of coordination shells number $m$ on the prediction accuracy. Fig. \ref{fig:s33_1} - Fig. \ref{fig:s33_3} show the BIC values for the different numbers of shells given various sizes of the training dataset. Using ensemble sampling strategy, we randomly collect data (from 20 data to 200 data) from each of four supercells by 100 random trials and then estimate the mean and standard deviation of BIC values for each specific number of shells. For NbTaMoW in Fig. \ref{fig:s33_1} (a), $m=2$ corresponding to the smallest BIC value represents the ``best" number of shells given only 20 data, while a larger number of $m$ leads to the issue of overfitting. As more data are collected into the model, the best number of shells gradually increases and converges towards $m=9$. When relatively large data ($n_d=150$) is available, the mean of BIC values with $m>9$ are quite close and the variation of BIC in each $m$ is much smaller than the case of small dataset size. NbTaMoWV as shown in Fig. \ref{fig:s33_2} has a similar overall trend with NbMoTaW but $m=5$ is identified for the case of $n_d<75$. NbTaMoWTi also tends a smaller number of shells if only limited data is provided but it displays an early converged number of shells $m=6$ other than NbTaMoW and NbTaMoWV that are $m=9$. 

Table \ref{tab:t2} provides a complete list of best number of shells given a specific dataset size (from 20 to 200 of each supercell) for these three refractory HEAs. If we divide the data into three categories, such as small size (total number of data $n_t = 4\times n_d$, $n_t<100$), medium size ($100 < n_t <400$) and large size ($n_t>400$), we can briefly conclude as follows:

\begin{itemize}
\item Small size: select a small number of shells $m=2$ or $m=3$ avoid overfitting
\item Medium size: identify $m=5$ or $m=6$ as a reliable choice avoid underfitting 
\item Large size: determine the reasonable large number of feature $6 \le m \le 9$ due to the bias-variance tradeoff \cite{mehta2019high}
\end{itemize}

\begin{table}[!ht] \footnotesize
\centering
\caption{The best number of coordination shells given a specific size of training dataset}
\label{tab:t2}
\begin{tabular}{@{}cccc@{}}
\toprule
Each supercell & NbMoTaW & NbMoTaWV & NbMoTaWTi \\ \midrule
20  & 2   & 3   & 2   \\
25  & 2   & 3   & 3   \\
30  & 5   & 5   & 3   \\
35  & 5   & 5   & 5   \\
40  & 5   & 5   & 5   \\
45  & 5   & 5   & 6   \\
50  & 6   & 5   & 6   \\
55  & 6   & 5   & 6   \\
60  & 6   & 5   & 6   \\
65  & 6   & 5   & 6   \\
70  & 9   & 5   & 6   \\
75  & 9   & 6   & 6   \\
100 & 9   & 9   & 6   \\
125 & 9   & 9   & 6   \\
150 & 9   & 9   & 6   \\
175 & 9   & 9   & 6   \\
200 & 9   & 9   & 6   \\ \bottomrule
\end{tabular}
\end{table}

\begin{figure}[!ht]   
  \centering
  \subfigure[]{\includegraphics[width=0.4\textwidth]{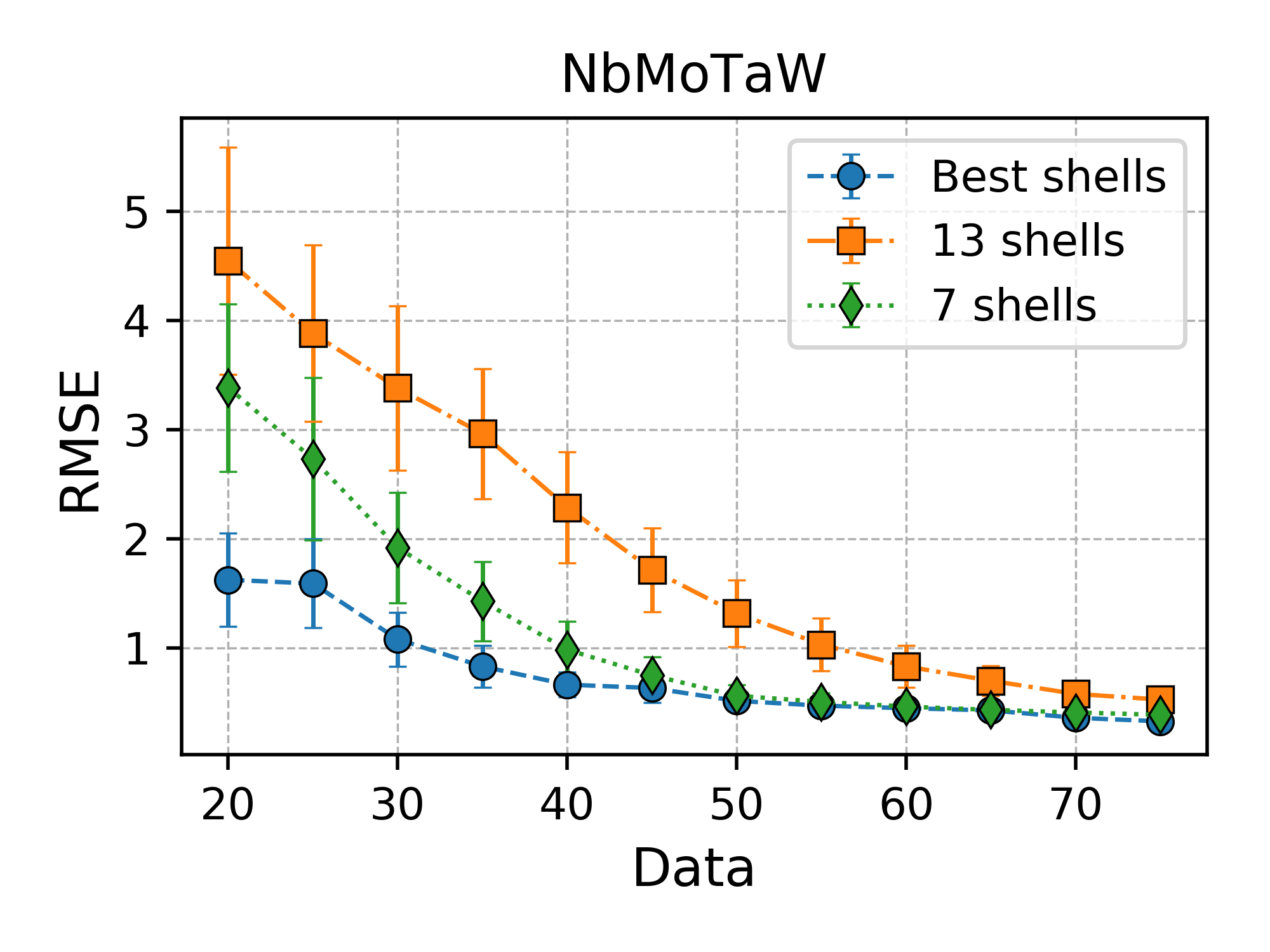}}
  \subfigure[]{\includegraphics[width=0.4\textwidth]{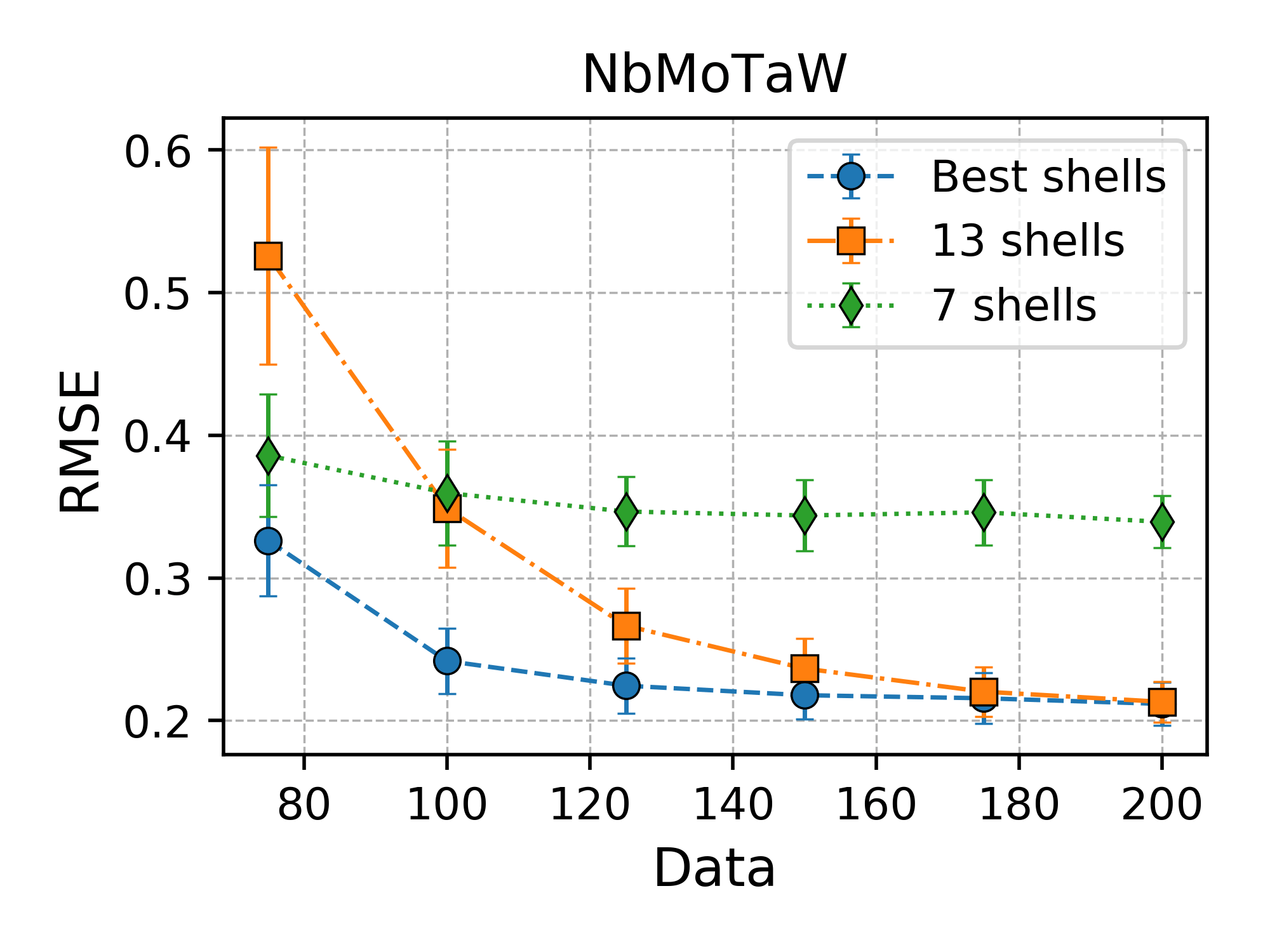}}
  \caption{RMSE results of NbMoTaW with three selections of coordination shells number given different sizes of training dataset (a) $n_d \le 75$ and (b) $n_d \ge 75$ } \label{fig:s33_4}
\end{figure}

Next, let's turn our attention to investigate the effect of feature selection on the testing RMSE performance. As shown in Fig. \ref{fig:s33_4}, we compare the RMSE results with three numbers of shells that include $m=13$, $m=7$ and $m=m_{\textup{best}}$ where $m_{\textup{best}}$ is referred to as the best number of shells. It is easy to observe that the RMSE of $m=13$ ($\varepsilon_R = 4.5$ meV), $m=7$ ($\varepsilon_R = 3.4$ meV) are significantly larger than the results using the best number of shells that is only approximate $\varepsilon_R = 1.5$ meV given $n_d=20$ training data. The RMSE mean value of $m=13$ and $m=7$ is reduced to approximate the best shells as dataset size increases (see Fig. \ref{fig:s33_4} (a)) but the variations (standard deviation) are still substantially large. When a relatively large data is collected, as shown in Fig. \ref{fig:s33_4} (b) that has different y-axis scale from Fig. \ref{fig:s33_4} (a)), the RMSE of $m=13$ and $m=m_{\textup{best}}$ gradually converge to a small value, $\varepsilon_R = 0.21$ meV, while $m=7$ converges to $\varepsilon_R = 0.35$ meV and can not be further decreased. In other words, the shorter cutoff (for example $m=7$) for the physical feature does not fully capture all the physical information and underestimates the system complexity of HEAs. 

\begin{figure}[!ht]   
  \centering
  \subfigure[]{\includegraphics[width=0.4\textwidth]{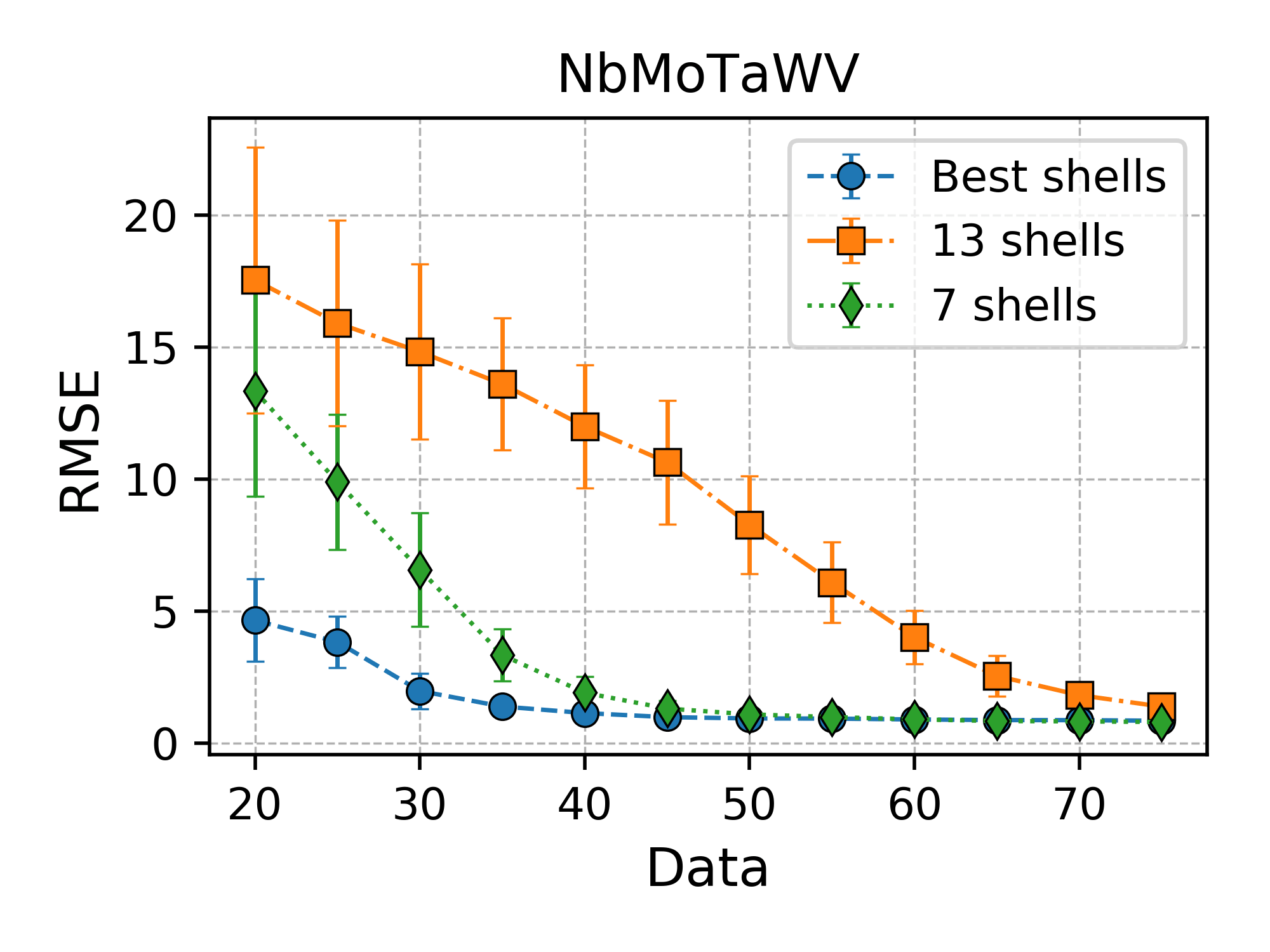}}
  \subfigure[]{\includegraphics[width=0.4\textwidth]{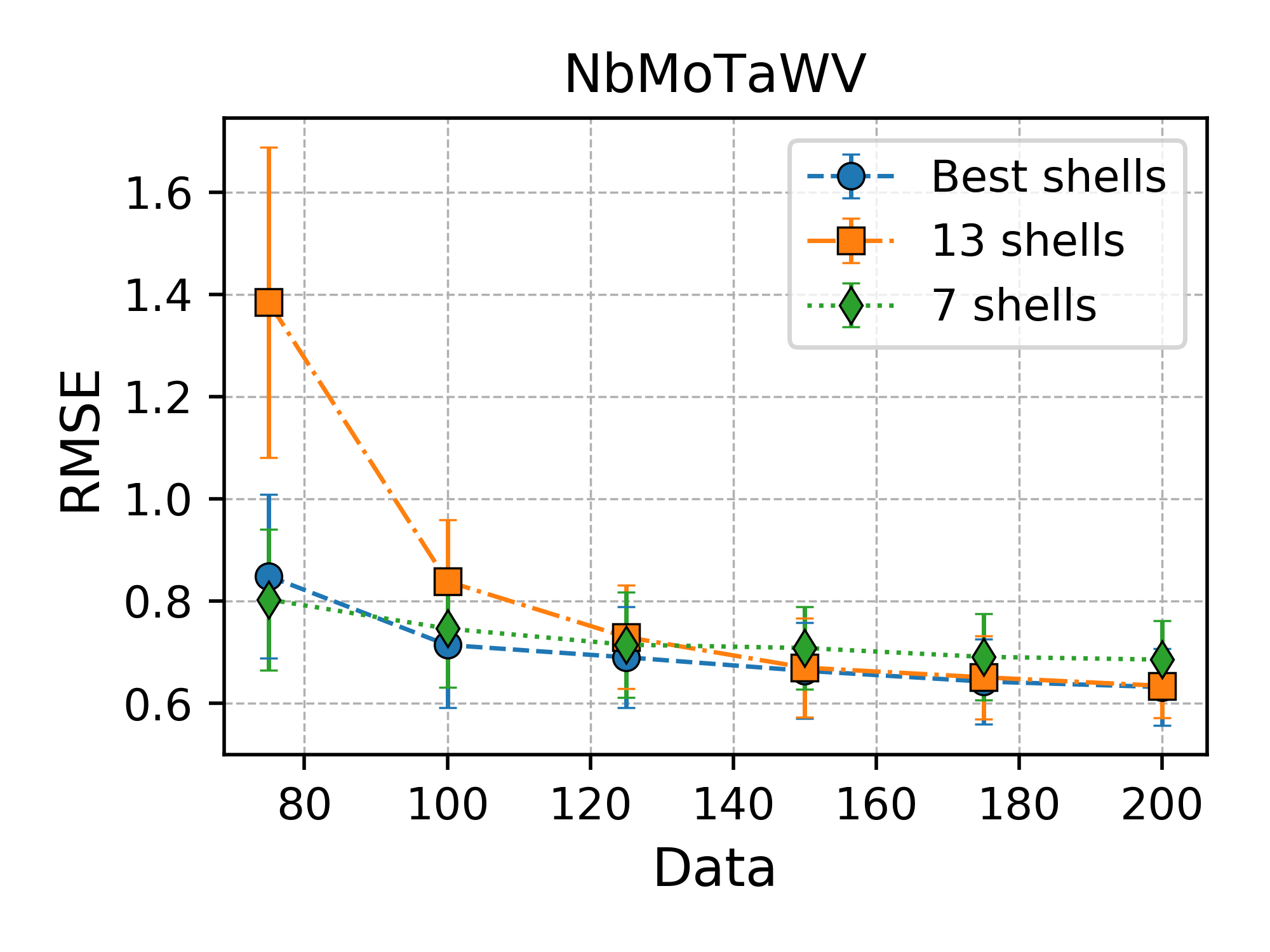}}
  \caption{RMSE results of NbMoTaWV with three selections of coordination shells number given different sizes of training dataset (a) $n_d \le 75$ and (b) $n_d \ge 75$ } \label{fig:s33_5}
\end{figure}

\begin{figure}[!ht]   
  \centering
  \subfigure[]{\includegraphics[width=0.4\textwidth]{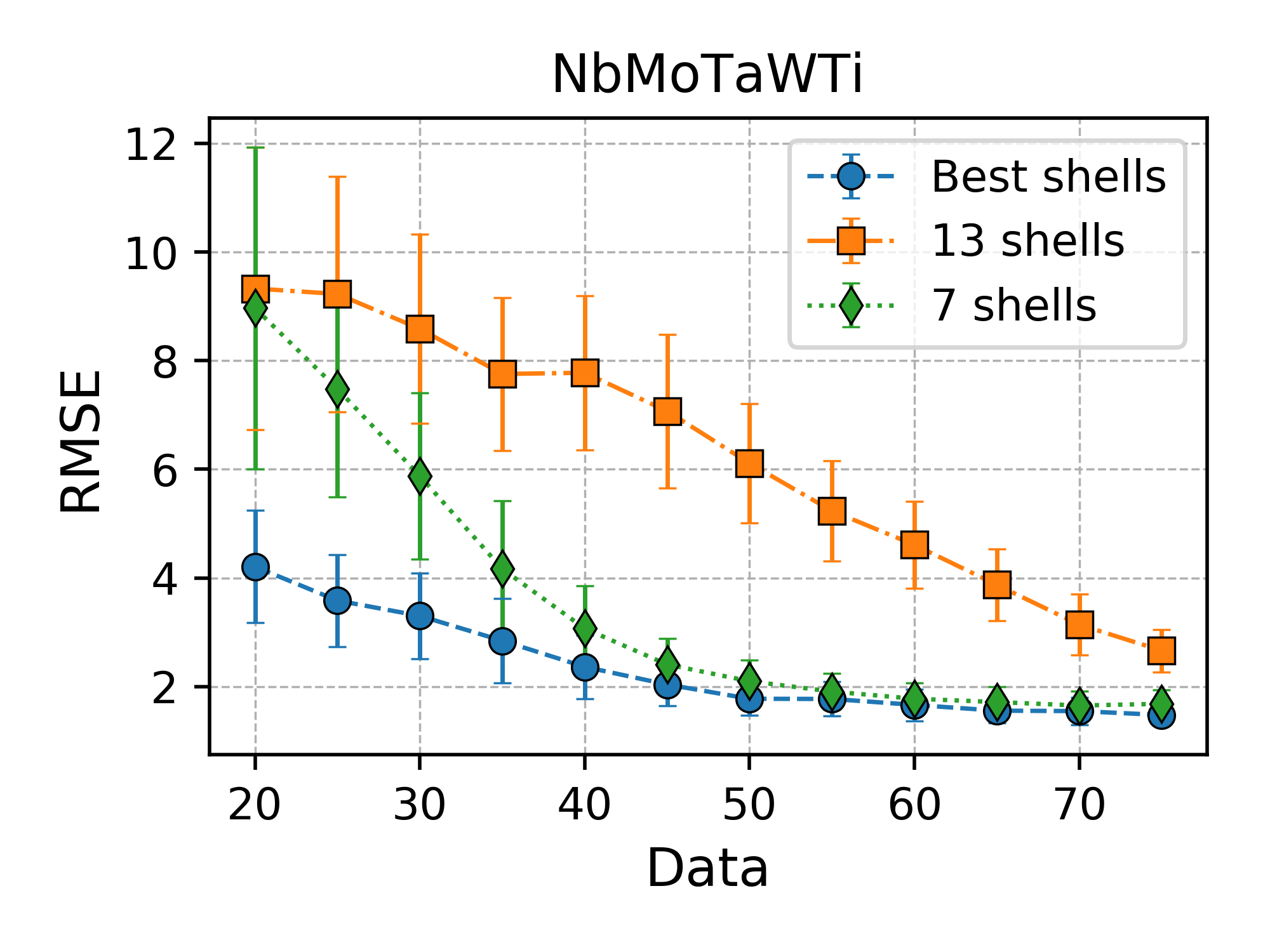}}
  \subfigure[]{\includegraphics[width=0.4\textwidth]{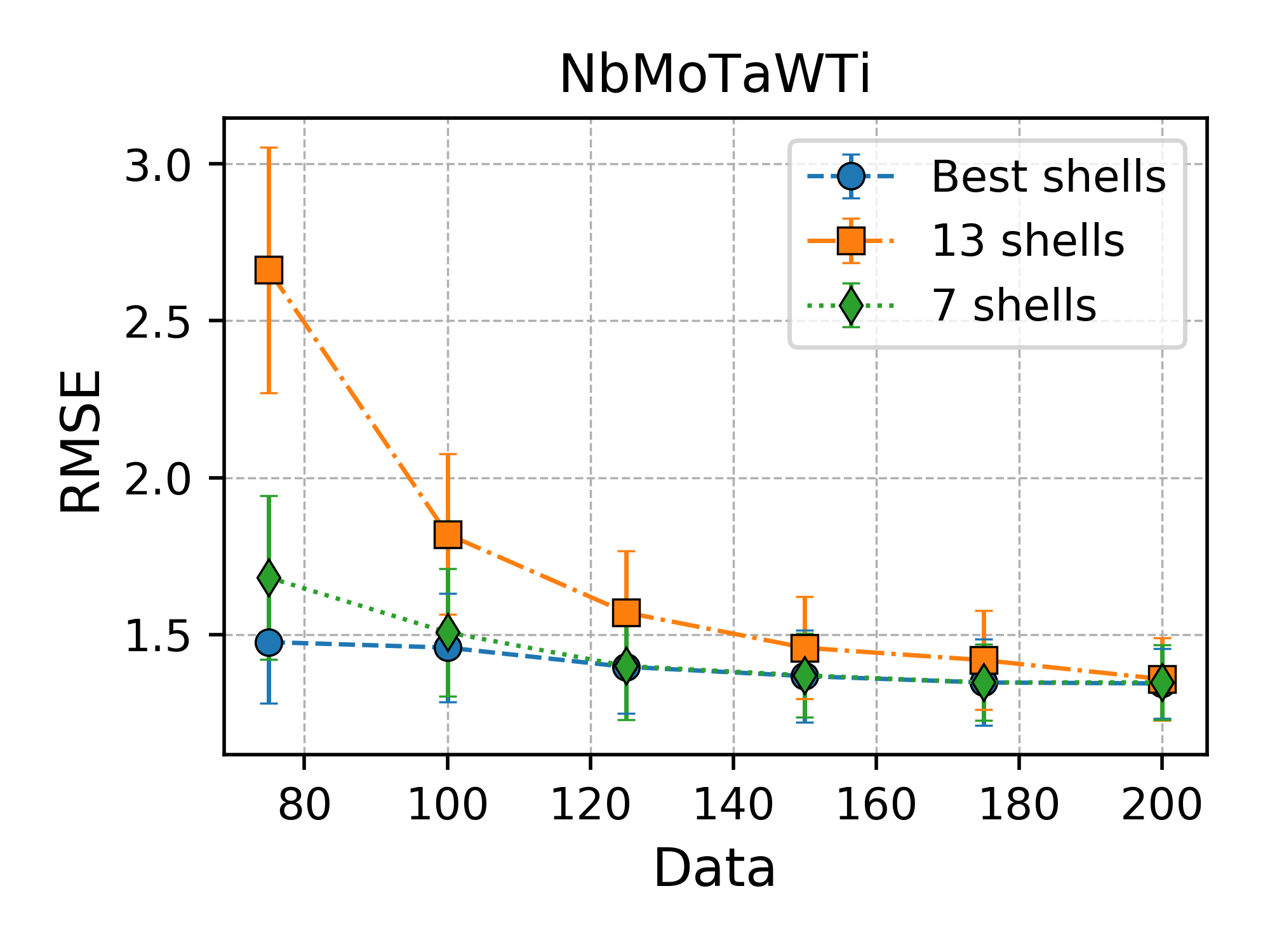}}
  \caption{RMSE results of NbMoTaWTi with three selections of coordination shells number given different sizes of training dataset (a) $n_d \le 75$ and (b) $n_d \ge 75$ } \label{fig:s33_6}
\end{figure}

Fig. \ref{fig:s33_5} and Fig. \ref{fig:s33_6} show the RMSE results for NbMoTaWV and NbMoTaWTi respectively. Given a small size of data ($n_d< 25$), the RMSE of $m=13$ and $m=7$ are nearly more than $\varepsilon_R = 10$ meV, which may lead to a large bias for Monte Carlo simulation of thermodynamics. Nevertheless, the best shells with the same data show a relatively small ($\varepsilon_R <5$ meV) and reliable RMSE estimate. In the medium size of data, the RMSE of $m=13$ and $m=7$ are reduced but still larger than that of the best shells, which can be observed from Fig. \ref{fig:s33_5} (a) and Fig. \ref{fig:s33_6}(a). When relatively large data is considered, $m=13$ and $m=7$ in NbMoTaWTi eventually converge towards the minimal value as the best shells, while $m=7$ in NbMoTaWV still shows slight underfitting issue because the best number identified by BIC is $m=9$ greater than the arbitrary choice of $m=7$. Through careful analysis and comparison of these three HEAs, we found that the model with the best number of shells identified by feature selection demonstrates a highly accurate and robust performance on either small or relatively large data. Rather than an arbitrary selection of physical feature, a reliable feature selection can effectively reduce the risk of underfitting and overfitting during the prediction. 

\section{Conclusions}

In this work, we develop a systematically robust Bayesian framework to discover efficient and accurate modeling of configurational energy from the data-driven perspective. A short-range pair interaction model is used here to characterize the physical feature and well-suited to deal with a large number of components inherent with the multicomponent systems. Bayesian regression algorithm is employed to establish an efficient Hamiltonian through a set of random configurations with the corresponding energy calculated by DFT method and to quantify the uncertainties and correlations of effective pair interaction parameters. To improve the accuracy and reliability of prediction, we further perform Bayesian feature selection for dealing with the truncation of the model, specifically given lack of data. All three HEAs, NbTaMoW, NbTaMoWV and NbTaMoWTi have demonstrated a highly accurate and robust performance in predicting the configurational energy. The proposed method is therefore a powerful tool for studying the thermodynamics and order-disorder phase transitions through the subsequent Monte Carlo simulations. 

Specifically, we find that a small and single supercell is unable to well explore the various order and disorder in multicomponent systems. We therefore propose an ensemble sampling strategy which naturally incorporates chemical configurations of different short-range and long-range order, thus performs a well-suited sampling capability of the huge configuration space so that it has demonstrated to be a simple yet robust scheme to enhance the representativity of data. Using this strategy, the resulted RMSE of three HEAs show a very small mean value ($\varepsilon <1 meV$) with a tiny standard deviation that is significantly lower than the other cases that use only single supercell. 

Also, we note that the chemical bonds in the EPI model show a frustrating behavior if the dataset size is too small ($n_t=100$). This can also be observed from the uncertainty quantification of EPI bonds, for instance, the nearest-neighbor shell displays a high level. However, as more data are obtained, for example, from $n_t=100$ to $n_t=400$, the bonds tend to be stable and the uncertainties are reduced. Moreover, we find a clear pattern from the variance-covariance matrix that characterizes each coordination shell as the individual physical feature consisting of several sub-features that depends on the number of component species. We also notice a certain degree of correlation within the nearest-neighbor shell even though they are essentially assumed to be independent. 

Finally, the impact of feature selection and the effect of dataset size are carefully discussed. For each material, the best truncated number of coordination shell is slightly different given a specific dataset but they demonstrate a similar trend. We therefore provide a general suggestion according to the dataset size: a) $m=2 \sim 3$ for small size ($n_t<100$), $m=5 \sim 6$ for medium size ($100 \le n_t \le 400$) and $m=6 \sim 9$ for relatively large size ($n_t>400$). Using this feature selection scheme, we have demonstrated an accurate and robust performance for all three HEAs without concerning the issue of underfitting or overfitting that is often happened in machine learning modeling. 

\section{Acknowledgements}
This work of J. Z. was supported by the Laboratory Directed Research and Development Program of Oak Ridge National Laboratory. X. L. and M. E. were supported by the U.S. Department of Energy, Office of Science, Basic Energy Sciences, Materials Science and Engineering Division. This research used resources of the Oak Ridge Leadership Computing Facility, which is supported by the Office of Science of the U.S. Department of Energy under Contract No. DE-AC05-00OR22725.

\bibliographystyle{model1-num-names}
\bibliography{sample.bib}







\end{spacing}
\end{document}